\documentclass{aastex631}
\usepackage[utf8]{inputenc}
\usepackage{graphicx}
\usepackage{txfonts}
\usepackage[shortlabels]{enumitem}
\usepackage{ulem}

\accepted{June, 6, 2024}

\submitjournal{ApJ}

\graphicspath{{./}{}}

\begin{document}

\title{The relative prevalence of wave-packets and coherent structures in the inertial and kinetic ranges of turbulence as seen by Solar Orbiter}

\author{Alina Bendt}
\affiliation{Centre for Fusion, Space And Astrophysics, Physics Department, University of Warwick, UK}

\author[0000-0003-0053-1584]{Sandra Chapman}
\affiliation{Centre for Fusion, Space And Astrophysics, Physics Department, University of Warwick, UK}
\affiliation{International Space Science Institute, Bern, Switzerland}
\affiliation{Department of Mathematics and Statistics, University of Tromso, Norway}

\author{Thierry Dudok de Wit}
\affiliation{International Space Science Institute, Bern, Switzerland}
\affiliation{University of Orléans}

\begin{abstract}
The Solar Orbiter (SO) mission provides the opportunity to study the evolution of solar wind turbulence. 
We use SO observations of nine extended intervals of homogeneous turbulence to determine when turbulent magnetic field fluctuations may be characterized as: (i) wave-packets and (ii) coherent structures (CS). 
We perform the first systematic scale-by-scale decomposition of the magnetic field using two wavelets known to resolve wave-packets and discontinuities, the Daubechies 10 (Db10) and Haar respectively. 
The probability distributions (pdfs) of turbulent fluctuations on small scales exhibit stretched tails, becoming Gaussian at the outer scale of the cascade.
Using quantile-quantile plots, we directly compare the wavelet fluctuations pdfs, revealing three distinct regimes of behaviour.
Deep within the inertial range (IR) both decompositions give essentially the same fluctuation pdfs.
Deep within the kinetic range (KR) the pdfs are distinct as the Haar decompositions have larger variance and more extended tails.
On intermediate scales, spanning the IR-KR break, the pdf is composed of two populations: a core of common functional form containing $\sim97$\% of fluctuations, and tails which are more extended for the Haar decompositions than the Db10 decompositions.
This establishes a crossover between wave-packet (core) and CS (tail) phenomenology in the IR and KR respectively.
The range of scales where the pdfs are $2$-component is narrow at $0.9$ au ($4-16$ s) and broader ($0.5-8$ s) at $0.4$ au.
As CS and wave-wave interactions are both candidates to mediate the turbulent cascade, these results offer new insights into the distinct physics of the IR and KR.
\end{abstract}


\section{Introduction}\label{sec:intro}
The super Alfvenic, high Reynolds number solar wind flow provides a large scale natural laboratory for plasma turbulence (see e.g. \cite{tu_mhd_1995, bruno_solar_2013, chen_recent_2016, marino_scaling_2023}).
There are extensive observations at $1$ au (e.g. from ACE, WIND and Cluster) principally around the $L1$ point upstream of earth (for a review see e.g. \cite{bruno_solar_2013, verscharen_multi-scale_2019}).
Previous observations at different distances from the sun have been provided by e.g. Helios, Ulysses and Voyager (see e.g. \cite{bruno_solar_2013, nicol_signature_2008, cuesta_intermittency_2022, yordanova_turbulence_2009, bourouaine_spectral_2012, maruca_trans-heliospheric_2023, pagel_radial_2003, bavassano_statistical_1982, tu_power_1984, roberts_heliocentric_1990}).
Solar Orbiter \citep{muller_solar_2013, muller_solar_2020} and Parker Solar Probe offer new opportunities to study the solar wind at different distances from the sun from $1$ au to within $0.1$ au.

Results around $1$ au consistently show features of turbulence phenomenology. 
The power spectrum of magnetic field fluctuations in the trace and components exhibits a well defined inertial range (IR) of magneto-hydrodynamic (MHD) turbulence with a steeper kinetic range (KR) scaling below ion scales and a shallower, approximately $1/f$-range at larger scales (e.g. \cite{kiyani_dissipation_2015}).
The IR trace power spectrum typically exhibits a power spectral scaling around $-5/3$ (e.g. \cite{matthaeus_measurement_1982, beresnyak_basic_2012, podesta_spectral_2007}), which corresponds to the Kolmogorov 1941 (K41) scaling \citep{kolmogorov_local_1997}.
Closer to the sun, at distances smaller than $0.4$ au (e.g. \cite{safrankova_evolution_2023, chen_evolution_2020, lotz_radial_2023}) the power spectrum on average evolves towards a spectral slope of $-3/2$, which corresponds to Iroshnikov-Kraichnan (IK) scaling \citep{iroshnikov_turbulence_1963}.
Below ion kinetic scales the spectrum steepens to a well defined kinetic range (e.g. \cite{sahraoui_evidence_2009, chen_ion-scale_2014, verscharen_multi-scale_2019, kiyani_enhanced_2013}).
The steeper kinetic range power spectrum corresponds to an increase in compressibility \citep{kiyani_global_2009, kiyani_enhanced_2013, alexandrova_small-scale_2008, alexandrova_solar_2013} compared to the IR.

Both waves and coherent structures are features of MHD turbulent phenomenology \citep{tu_mhd_1995, frisch_turbulence_1995} and may mediate the turbulent cascade.
In the following, a coherent structure is a sudden discontinuity that stands out of the fluctuations. Recent studies of the KR reveal whistler waves, ion-cyclotron waves, and kinetic Alfvén waves as well as coherent structures in this regime (e.g. \cite{roberts_variability_2017, sahraoui_evidence_2009, wu_intermittent_2013, zhou_electron_2023, alexandrova_solar_2013, chhiber_subproton-scale_2021, osman_kinetic_2012, he_oblique_2011, salem_identification_2012}), where kinetic effects and ultimately dissipation become important (e.g. \cite{kiyani_dissipation_2015, verscharen_multi-scale_2019}).
A feature of turbulence, is intermittency, which has been identified by \cite{koga_intermittent_2007} as arising from phase correlation among different scales due to nonlinear wave-wave interactions and as coherent structures by \cite{gomes_origin_2022, camussi_orthonormal_1997} and \cite{veltri_mhd_1999}. These coherent structures have also been identified as localized sites of turbulent dissipation \citep{perri_detection_2012, greco_partial_2017, wu_intermittent_2013, osman_kinetic_2012, osman_magnetic_2014, osman_intermittency_2012, sioulas_statistical_2022}.

Identification of turbulence rests upon statistical characterization, since quantitative aspects of turbulence are reproducible in a statistical sense and each realization is distinct \citep{frisch_turbulence_1995, tu_mhd_1995}. 
A key characteristic of turbulence is scale-by-scale similarity \citep{frisch_turbulence_1995, tu_mhd_1995}. The process of statistical characterization and testing for scaling includes a two-step process: (i) obtain the fluctuation time-series decomposition, by differencing, Fourier \citep{welch_use_1967} or wavelet decomposition \citep{farge_wavelet_1991, meneveau_analysis_1991, daubechies_wavelet_1990, mallat_theory_1989}, followed by (ii) analyse the fluctuations scale by scale by examining power spectra and pdfs.
All the above methods are in widespread use in the study of solar wind turbulence (eg. \cite{podesta_spectral_2007, kiyani_enhanced_2013, camussi_orthonormal_1997, farge_wavelet_1992, yamada_orthonormal_1991, do-khac_wavelet_1994, narasimha_wavelet_2007, bolzan_comparisons_2009, beresnyak_basic_2012, bruno_solar_2013, chapman_quantifying_2007, katul_estimating_2001}).

Turbulent fluctuations in solar wind data extracted by differencing the time-series have non-Gaussian probability distributions (pdfs)
\citep{bruno_solar_2013, frisch_turbulence_1995, tu_mhd_1995, alexandrova_small-scale_2008, bruno_probability_2004, bruno_radial_2003, sorriso-valvo_intermittency_1999, hnat_intermittency_2003}
which tend to become more Gaussian on scales approaching the outer scale of the turbulent cascade. 
The stretched exponential tails of the pdfs \citep{hnat_intermittency_2003}, hereafter referred to as stretched tails, show that large fluctuations have a higher probability of occurrence than for a Gaussian distribution, consistent with intermittency (\cite{bruno_intermittency_2019} and references therein).

In this paper, we use wavelet decompositions and perform the first systematic comparison of different features in magnetic field records by considering two types of mother wavelets: (i) Haar or first order Daubechies wavelets are well suited for capturing sharp discontinuities (hereafter called coherent structures) such as the signature of current sheets, while (ii) 10th order Daubechies wavelets have a wave-like shape, and therefore are better adapted for detecting wave packets, see Figure \ref{fig:wavelets}. By comparing the two wavelet decompositions we should then be able to distinguish the role of these different features at different scales of the turbulent cascade.
Different time-series decompositions extract different features in the time-series \citep{schneider_computing_2001, farge_wavelet_1991, farge_wavelet_1992}. 
We will see that comparing different decompositions of the time-series can identify how coherent structures and wave-wave interactions contribute to the turbulent cascade.

The IR of solar wind turbulence is anisotropic due to the presence of a background magnetic field \citep{matthaeus_evidence_1990}
as seen in the power spectrum (e.g. \cite{bruno_solar_2013, oughton_anisotropy_2015, bandyopadhyay_geometry_2021, chen_anisotropy_2011, horbury_anisotropic_2008, wicks_power_2010}).
The background field that is expected to order the anisotropy of the magnetic fluctuations can be defined globally, averaging across scales and time, or locally, scale-by-scale and varying in time (e.g. \cite{horbury_anisotropic_2008, beresnyak_basic_2012, chapman_quantifying_2007, duan_anisotropy_2021, kiyani_enhanced_2013, podesta_dependence_2009, turner_nonaxisymmetric_2012, zhang_three-dimensional_2022, yamada_identification_1991}). These analyses can yield a broad spread in values of the power spectral exponent \citep{tessein_spectral_2009} and differing estimates of its anisotropy \citep{oughton_anisotropy_2015}. In this paper we will consider the former, global background field.
Averaging the magnetic field vector over a global timescale exceeding that of the centre scale of the turbulence defines a global background field. Together with the time-averaged solar wind velocity, a coordinate system is constructed. The time average is typically taken over the entire intervals of data (in this study we use intervals from $10$ to $31.5$ h length) \citep{bruno_solar_2013}.

In this paper we will find that the IR-KR transition can, depending upon conditions, coincide with the crossover to a region where coherent structures dominate the population of large fluctuations.
By comparing different decompositions of the time-series in a global background field, we find that coherent structures are prevalent in the KR and less dominant in the IR.
The temporal scale where the PSD steepens from the IR to the KR is indicative of a transition from MHD to ion kinetic physics. There has been considerable effort to identify this scale break, and it does not necessarily appear at the same scale for any plasma conditions \citep{chen_ion-scale_2014, markovskii_statistical_2008, wang_ion-scale_2018, safrankova_evolution_2023}. 
Generally, the spectral break occurs between $0.02 - 4$ Hz \citep{markovskii_statistical_2008}.
Recently, \cite{safrankova_evolution_2023} found that the spectral break decreases with heliocentric distance from around $4$ Hz close to the sun to $0.1$ Hz around $1$ au.

This paper is organised in three sections. In section \ref{sec:data-methods} we present the data intervals analysed and data analysis methods. In section \ref{sec:results} we present a systematic comparison of power spectra and fluctuation pdfs applied to two different scale-by-scale decompositions of the data.
We conclude in section \ref{sec:conclude}.

\section{Data and Methods}\label{sec:data-methods}
\subsection{Data}
We analyse in detail the time-series of magnetic field data from the Magnetometer (MAG) \citep{horbury_solar_2020} and obtain averaged parameters from the solar wind velocity, density, pressure and temperature measurements of the Solar Wind Analyser (SWA-PAS) \citep{owen_solar_2020} on board Solar Orbiter \citep{muller_solar_2013}. The solar wind velocity and magnetic field measurements are provided in $RTN$ coordinates, with the magnetic field measurements at a cadence of $8$ Hz.
We select nine over $10$ h long intervals of turbulence which contain homogeneous solar wind flow without any shocks, current sheet crossings and other large events, at heliocentric distances $R$ of $\sim 0.3, 0.6$, and $\sim 0.9$ au. Three intervals have a plasma $\beta\geq 1.7$.
The average solar wind velocity of the intervals, $V_{sw}$ is $494$ km s$^{-1}$. 
Table \ref{tab:scales} presents the intervals, grouped in four categories: i) the high plasma beta of $\beta\geq 2$ intervals from 2021-11-18 at $0.9$ au and 2023-03-14 at $0.6$ au (italics), ii) this encompasses the interval with a large field alignment angle $\theta$ (bold) iii) intervals close to the sun (double underlined) and, iv) intervals at $\sim 0.9$ au with moderate plasma $\beta$ (underlined).
We rotate the magnetic field from $RTN$ coordinates into coordinates ordered by the global time-averaged background field, averaged over the entire interval $B_{0}$. The orthogonal coordinate system then has the magnetic field projected onto a component $B_{\parallel}$ parallel to $B_{0}$, and onto perpendicular components $B_{\perp (V_{sw}, B)} = B_{\parallel} \times V_{sw}$, and $B_{\perp (V_{sw}\times B)}= B_{\parallel} \times B_{\perp (V_{sw}, B)}$.
\begin{table}[ht]
    \centering
    \begin{tabular}{|c|c|c|c|c|c|c|c|c|}
    \centering
        \textbf{interval} [Y-M-D] & \textbf{length} [h] & $R$ [au] & $\mathbf{V_{sw}}$ [km s$^{-1}$] & $\mathbf{\tau_{adv}}$ [h] & $\mathbf{\beta}$ & $\mathbf{\rho_{i}}$ [Hz] & \textbf{KR break} [Hz] & $\mathbf{\theta}$ [°] \\
        \underline{2022-01-01} & $\sim 14$ h & $0.997$ au & $584$ km s$^{-1}$ & $70.96$ h & $1.36$ & $0.31$ Hz & $0.5$ Hz & $27.14$° \\
        \underline{2022-01-03} & $\sim 15.3$ h & $0.992$ au & $530$ km s$^{-1}$ & $77.68$ h & $1.55$ & $0.19$ Hz & $0.5$ Hz & $31.82$° \\
        \underline{2022-01-04} & $\sim 10.75$ h & $0.989$ au & $438$ km s$^{-1}$ & $93.87$ h & $0.95$ & $0.24$ Hz & $0.25$ Hz & $18.07$° \\
        \underline{2022-01-06} & $\sim 24$ h & $0.984$ au & $312$ km s$^{-1}$ & $130.94$ h & $1.79$ & $0.23$ Hz & $0.5$ Hz & $64.51$° \\
        \textit{\textbf{2021-11-18}} & $\sim 14.75$ h & $0.934$ au & $533$ km s$^{-1}$ & $72.82$ h & $2.08$ & $0.19$ Hz & $1$ Hz & $160.93$° \\
        \textit{2023-03-14} & $\sim 10$ h & $0.597$ au & $548$ km s$^{-1}$ & $45.27$ h & $2.48$ & $0.68$ Hz & $1$ Hz & $68.63$° \\
        \uuline{2022-03-18} & $\sim 12$ h & $0.369$ au & $414$ km s$^{-1}$ & $37.08$ h & $0.98$ & $1.35$ Hz & $1$ Hz & $7.29$° \\
        \uuline{2022-04-04} & $\sim 24$ h & $0.369$ au & $555$ km s$^{-1}$ & $27.64$ h & $0.76$ & $1.22$ Hz & $1$ Hz & $16.82$°\\
        \uuline{2022-04-01} & $\sim 31.5$ h & $0.344$ au & $535$ km s$^{-1}$ & $26.66$ h & $1.02$ & $1.75$ Hz & $1$ Hz & $21.46$° \\
    \end{tabular}
    \caption{Table of the nine interval characteristics: the date, length, heliocentric distance $R$, average solar wind speed $V_{sw}$, advection times $\tau_{adv}$, plasma $\beta$, ion-gyro frequency $\rho_{i}$, the KR-IR spectral break time scale and the field alignment angle $\theta$. $R$ is quoted to three significant figures to distinguish intervals very close to each other, while the other parameters (except $V_{sw}$) are quoted to two decimal places. The spectral break scale is quoted for the perpendicular field components.}
    \label{tab:scales}
\end{table}

\subsection{Wavelet decompositions of the time-series and intermittency measures}
\begin{figure}
    \centering
    \includegraphics[width=0.9\linewidth]{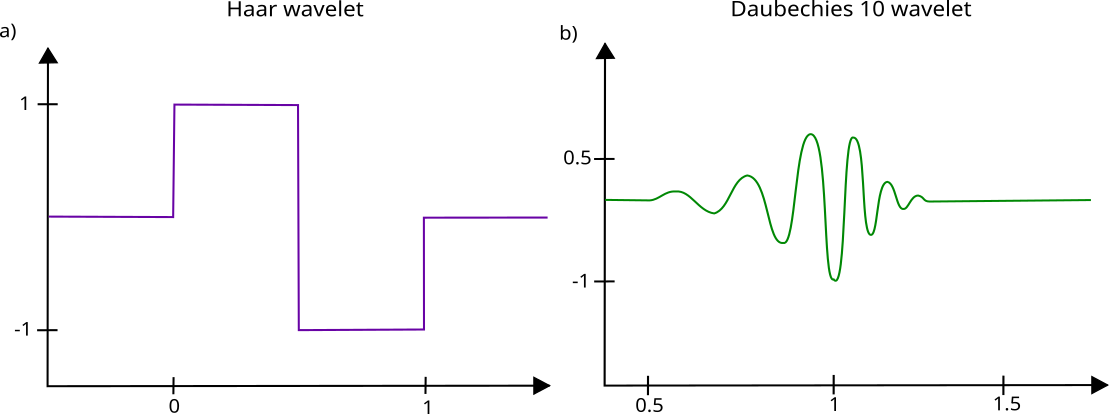}
    \caption{A schematic drawing of the Haar wavelet (a) and 10th order Daubechies wavelet (b) as functions of time.}
    \label{fig:wavelets}
\end{figure}

We decompose the magnetic field time-series of these nine intervals of homogeneous turbulence using two different discrete wavelet transforms, the 10th order Daubechies (Db10) and Haar wavelet (the latter is equivalent to differencing of the time-series). The different wavelets are designed to resolve wave-like features and sharp changes in the time-series respectively \citep{farge_wavelet_1992, percival_wavelet_2000, torrence_practical_1998, daubechies_wavelet_1990}.
A schematic drawing of the shape of the Haar and Db10 wavelets are presented in Figure \ref{fig:wavelets}.
Fourier, wavelet and differencing (structure functions) have all been used extensively in the study of solar wind turbulence, especially in testing for statistical scaling (e.g. \cite{podesta_spectral_2007, kiyani_enhanced_2013, farge_wavelet_1992, yamada_orthonormal_1991, do-khac_wavelet_1994, narasimha_wavelet_2007, bolzan_comparisons_2009, chapman_quantifying_2007, katul_estimating_2001}).
Wavelet decompositions are time-frequency localized and therefore are well suited to isolating wave-packets and coherent structures \citep{daubechies_wavelet_1990, farge_wavelets_1996}.
The discrete wavelet transform decomposes the time series into two components: a low-pass band, known as the approximation, and a high-pass band, known as the detail. The procedure is applied iteratively to the approximations which are on successively longer timescales. The cutoff frequencies define a dyadic sequence \citep{farge_wavelet_1992, percival_wavelet_2000} so that the set of wavelet details are the time-series band-passed filtered around central frequencies and bands of widths $2^{j} \Delta$, where $\Delta$ is the sampling period.

We will use $\tau$ to denote the scale of decomposition and $t_{k}$ as discrete time for the magnetic field time-series denoted as $B(t)$.
The wavelet details $\delta B_{\tau, t_{k}}$ at a time-scale $\tau = 2^{j} \Delta$, where $\Delta$ is the sampling period, and $j\in \mathbf{Z}$ the scale, and $t_{k}$ the location of the magnetic field $B(t)$ are \citep{farge_wavelets_1996}
\begin{equation}
    \delta B_{\tau, t_{j}} = \sum_{k=1}^{N}B(t_{k})\sqrt{\tau}\Psi (t_{k}-t_{j}) ~,
    \label{eq:details}
\end{equation}
where $N$ is the length of the data set and $\Psi_{j,i}$ is the set of wavelets.
The power spectrum can then be defined as \citep{farge_wavelet_1992, schneider_computing_2001}
\begin{equation}
    E(t', \tau) = \frac{2\Delta}{N}|\delta B_{\tau, t_{j}}|^{2}~.
\end{equation}

Wavelet transforms thus sample the frequency space logarithmically which is well suited to the determination of the power law exponent of the power spectrum \citep{mallat_theory_1989}.
The Haar wavelet is a first order Daubechies wavelet. The Daubechies family is defined from the base wavelet: \citep{nickolas_wavelets_2017}
\begin{equation}
    \Psi (x) = \sum_{k=0}^{2N-1} (-1)^{k-1}a_{k}\Phi(2x+k-1)
\end{equation}
and the scaling function $\Phi$:
\begin{equation}
    \Phi (x) = \sum_{k=0}^{2N-1} a_{k} \Phi (2x-k)
\end{equation}
where the coefficients $a_{k}$ have to satisfy several conditions, and $k\in \mathbb{Z}$ and $N\in \mathbb{N}$. The construction of the Daubechies family of wavelets can be found in \cite{nickolas_wavelets_2017}.
The Haar wavelet $H$ is a step-function $H_{j,k}(x)=2^{j/2}H(2^{j}x-k)$ \citep{nickolas_wavelets_2017}. Since the Haar wavelet shape corresponds to sharp changes it will be sensitive to coherent structures.
The 10th order Daubechies wavelet (Db10) is determined from a base wavelet with $10$ wavelet coefficients \citep{daubechies_ten_1992, percival_wavelet_2000} and its shape corresponds to that of wave-packets. The Db10 wavelet has a higher number of vanishing moments than the Haar wavelet, enabling a more accurate determination of steeper power law exponents in the kinetic range \citep{farge_wavelets_1996}.
Power spectral estimates rely upon an accurate estimation of the total power in each discrete frequency band, which can be distributed linearly (Fourier), or in steps of $2^{k}$ (wavelets). The fidelity of the power spectrum will depend upon the method used to obtain the time-series in each frequency band. It has been shown previously that for spectra steeper than $-3$ the Haar wavelet does not produce converging estimates for the power in each frequency band \citep{cho_simulations_2009} and therefore, cannot be used to estimate the spectral scaling steeper than $–3$. For that reason the wavelet transform with Haar mother wavelets should be interpreted with great care in the KR. However, both the Haar and Db10 wavelet decompositions are well-defined methods to resolve fluctuations on different temporal scales from a time-series \citep{nickolas_wavelets_2017, percival_wavelet_2000}. The Haar wavelet performs a differencing that is similar \citep{lovejoy_haar_2012} to that used in the well-known structure function  defined as $S_{q}(\tau)=\langle |\delta B(\tau, t_{j})|^{q} \rangle_{t} \sim \tau^{\zeta(q)}$, which are at the core of analysis of turbulent fields \citep{frisch_turbulence_1995, tu_mhd_1995}. Therefore, we use both Haar and Db10 wavelet decompositions to study the fluctuation pdfs and how they vary with temporal scale.
The wavelet transform is performed by the Maximum Overlap Discrete Wavelet Transform (MODWT) \citep{hess-nielsen_wavelets_1996, percival_wavelet_2000} with reflected boundaries.

\begin{figure}[ht]
    \centering
    \includegraphics[width=0.9\linewidth]{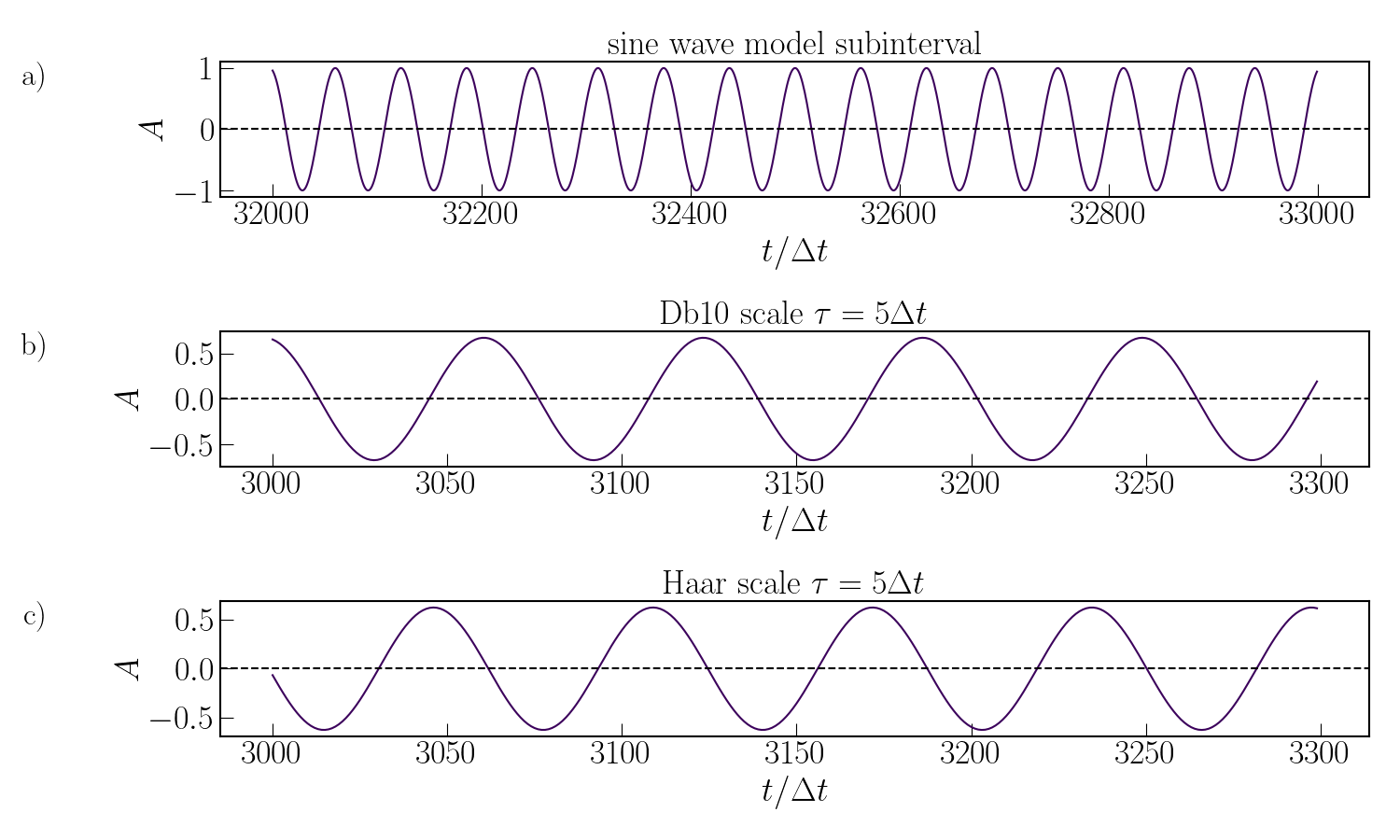}
    \caption{Wavelet decomposition of a sine wave model at  scale  $\tau=5\Delta t$ (a) using the Db10 wavelet (b) and the Haar wavelet (c). Both wavelet decompositions produce a sine wave with the same amplitude.}
    \label{fig:sine-decomposition}
\end{figure}
\begin{figure}[ht]
    \centering
    \includegraphics[width=0.9\linewidth]{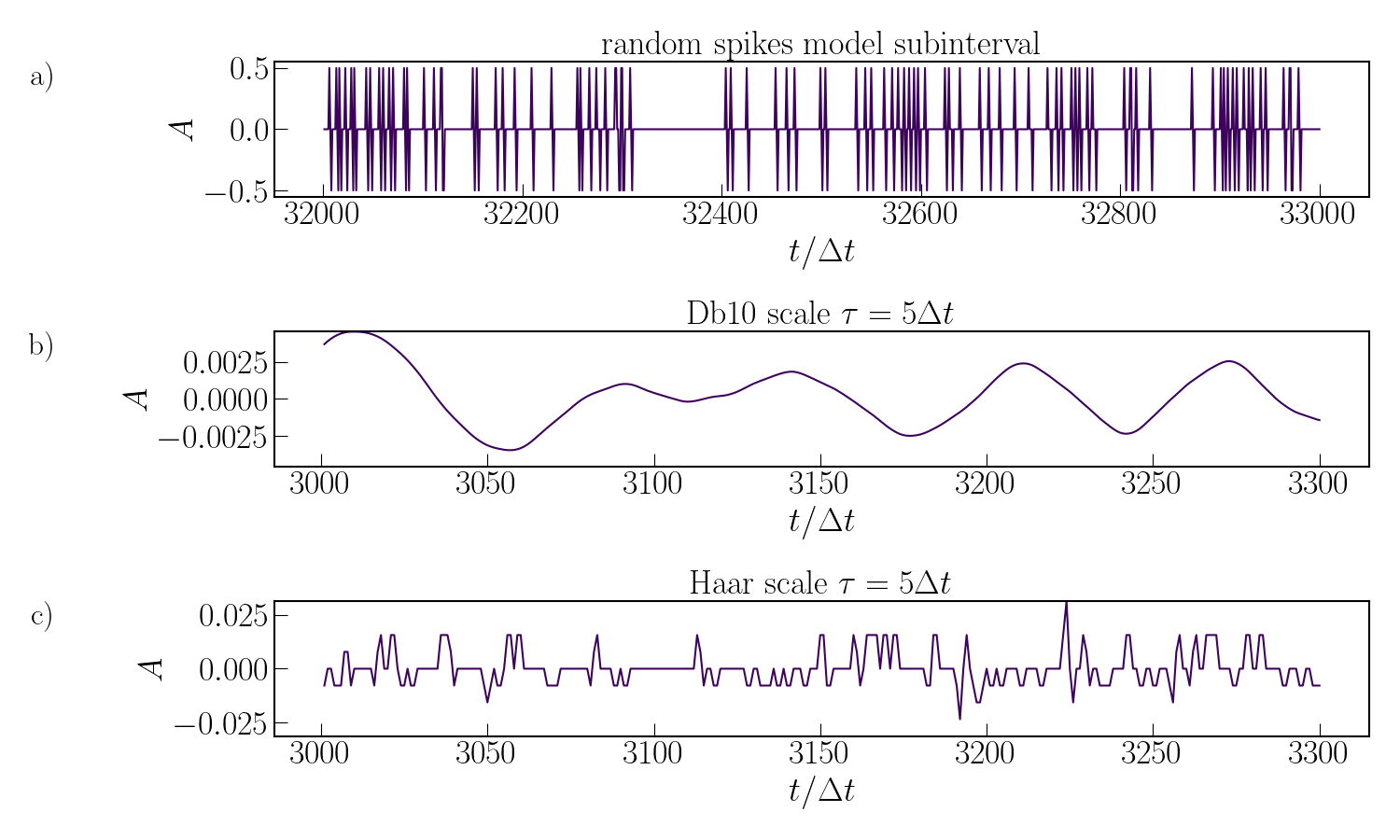}
    \caption{Wavelet decomposition of a random spike model at  scale $\tau=5\Delta t$  (a) using the Db10 wavelet (b) and the Haar wavelet (c). The Haar wavelet decomposition (c) more closely reflects the fine structure of the original time-series  compared to the Db10 wavelet decomposition (b). The Haar wavelet decomposition  amplitude is approximately an order of magnitude larger than the Db10 wavelet decomposition amplitude.}
    \label{fig:spike-decomposition}
\end{figure}
A simple illustration of the different performance of the Haar and Db10 wavelet decompositions is provided in Figures \ref{fig:sine-decomposition} and \ref{fig:spike-decomposition}. The wavelet decompositions are applied to a sine wave (Figure \ref{fig:sine-decomposition}) and a random spike train (Figure \ref{fig:spike-decomposition}).
The two wavelet decompositions perform similarly on the sine wave, in both cases a sinusoid of similar amplitude is resolved. The phase shift between panels (b) and (c) of Figure \ref{fig:sine-decomposition} is due to the different time asymmetry properties of the wavelets. For the random spike-train, the two wavelet decompositions yield quite distinct time-series, the Haar wavelet decomposition more closely reflects the time structure of the original time-series, and is about an order of magnitude larger in amplitude than the Db10 wavelet decomposition time-series. Comparing these two decompositions, and in particular the pdf of their coefficients we can then discriminate between time-series that contain a predominance of sharp discontinuities (i.e. coherent structures) and wave packets. Case (i) will yield fluctuations of similar amplitude, whereas case (ii) will yield Haar wavelet decomposed fluctuations of significantly larger amplitude than fluctuations obtained by the Db10 wavelet decomposition.

\section{Results}\label{sec:results}
We obtain scale-by-scale decompositions of the $9$ intervals using both Haar and Db10 wavelets, which then provide estimates of the power spectra and the fluctuation pdfs and their moments scale-by-scale. The aim is twofold: (i) to verify that the selected intervals do indeed exhibit properties consistent with turbulence phenomenology; (ii) by comparing the results of these analyses for the Haar (that is time-series differences) and the Db10 wavelets, to gain new insights into the relative importance of coherent structures and wave-like features at different temporal scales across the turbulent cascade.

\subsection{Power spectra}\label{subsec:establish}
\begin{figure}[ht]
    \centering
    \includegraphics[width=0.9\linewidth]{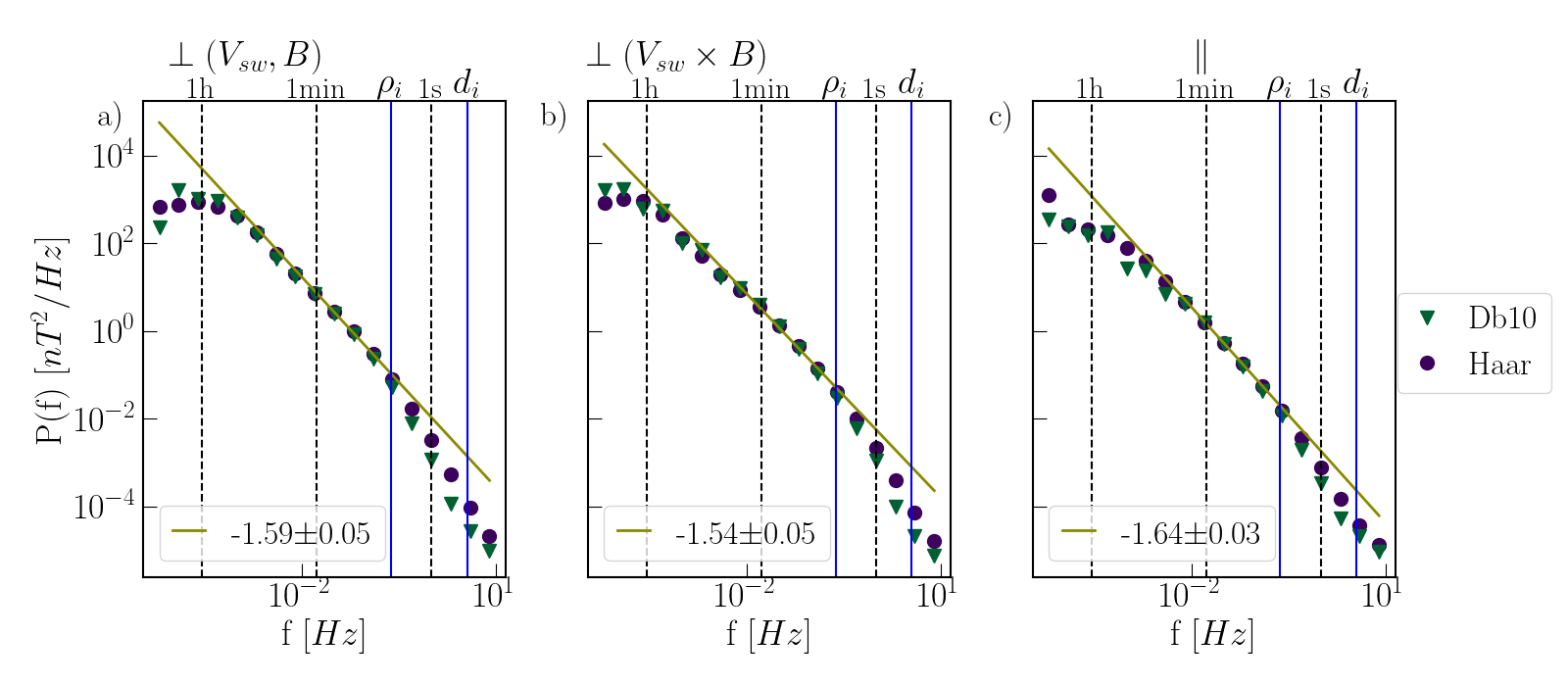}
    \caption{Power Spectral Density (PSD) with clear $1/f$, inertial and kinetic ranges in both Haar and Db10 wavelet decompositions. PSD are shown for one representative interval for each magnetic field component (column) and the respective scaling exponent in the inertial range fitted on the Haar wavelet decompositios. The interval is at $0.989$ au with $\beta=0.95$ and $\theta=18.07$° from 2022-01-04. The Db10 wavelet decomposition is shown by green triangles, purple circles are the Haar wavelet decomposition. Blue vertical lines denote the ion-gyro frequency $\rho_{i}$ and ion-inertia length $d_{i}$. Dashed black vertical lines denote scales marked on the x-axis as $1$ s, $2$ s, $1$ min, $1$ h. Yellow fit lines to the Haar wavelet power spectrum show the spectral exponent, which is quoted to three significant figures. }
    \label{fig:psd}
\end{figure}

We first establish that the power spectral estimates (Figure \ref{fig:psd}) of the Haar and Db10 discrete wavelets, show a clearly defined inertial range with power spectral breaks at low frequencies to the $1/f$-range and at high frequencies to the kinetic range, consistent with a well developed turbulence cascade.
Figure \ref{fig:psd} presents the power spectral density (PSD) for a representative interval for all magnetic field components, (a) $B_{\perp (V_{sw}, B)}$, (b) $B_{\perp (V_{sw}, B)}$, (c) $B_{\parallel}$. The full set of PSDs for all intervals is presented in Figure \ref{fig:psd-full}. Each spectrum is a single estimate using the full temporal range of each interval and is not averaged.
The parallel magnetic field component consistently shows less power than the perpendicular components (e.g. \cite{safrankova_evolution_2023}).
The intervals closer to the sun overall show more power at all scales (Figure \ref{fig:std} presents the standard deviation of the wavelet fluctuations for all intervals), as previously observed by \cite{chen_evolution_2020}.
The spectral exponents generally do not present clear IK or K41 scaling but rather values that lie between those values, as also seen by \cite{wang_wavelet_2023}.

As expected, the Haar and Db10 wavelet estimates diverge in the KR, seen in Figure \ref{fig:psd}, as the Haar cannot resolve scaling exponents steeper than -3 \citep{cho_simulations_2009}. Therefore, the Haar wavelet cannot reliably identify the scaling exponent in the KR, however it does provide appropriate fluctuations from the decomposition.
However, both the Haar and Db10 spectral estimates, within their given frequency resolution, identify the same location of the spectral break, which is identified as the smallest scale at which the wavelet PSDs coincide.
The IR-KR spectral break scale moves with the larger of the ion scales $\rho_{i}$ and $d_{i}$ (blue vertical lines in Figure \ref{fig:psd}, which is reproduced for all intervals in Figure \ref{fig:psd-full}), decreasing with decreasing distance from $4-1$ s and plasma $\beta ~\geq 2$. The evolution of the spectral break was previously observed by \cite{safrankova_evolution_2023, lotz_radial_2023, bruno_radial_2014} for magnetic field trace spectra.
For $B_{\parallel}$ the spectral break differs by one dyadic scale to the perpendicular components for $\beta \geq 2$ and the interval 2022-01-03 at $0.992$ au.

The outer inertial range spectral break to the $1/f$-range (Figure \ref{fig:psd}) is typically located before the $1$ h scale. The early $1/f$-break is most evident in $B_{\parallel}$ and $B_{\perp (V_{sw}, B)}$. The break between the IR and 1/f is well resolved in our wavelet spectral estimates which do not require multi-sample averaging, the break frequency  decreases with decreasing distance.
This was also reported by \cite{chen_evolution_2020}, who averaged each interval over a sliding window Fourier magnetic field trace spectra to 
obtain the break at $\sim 10^{4}$ s for large and $\sim 10^{3}$ s for small distances from the sun.

\subsection{Fluctuation PDFs scale by scale}
\begin{figure}[ht]
    \centering
    \includegraphics[width=0.9\linewidth]{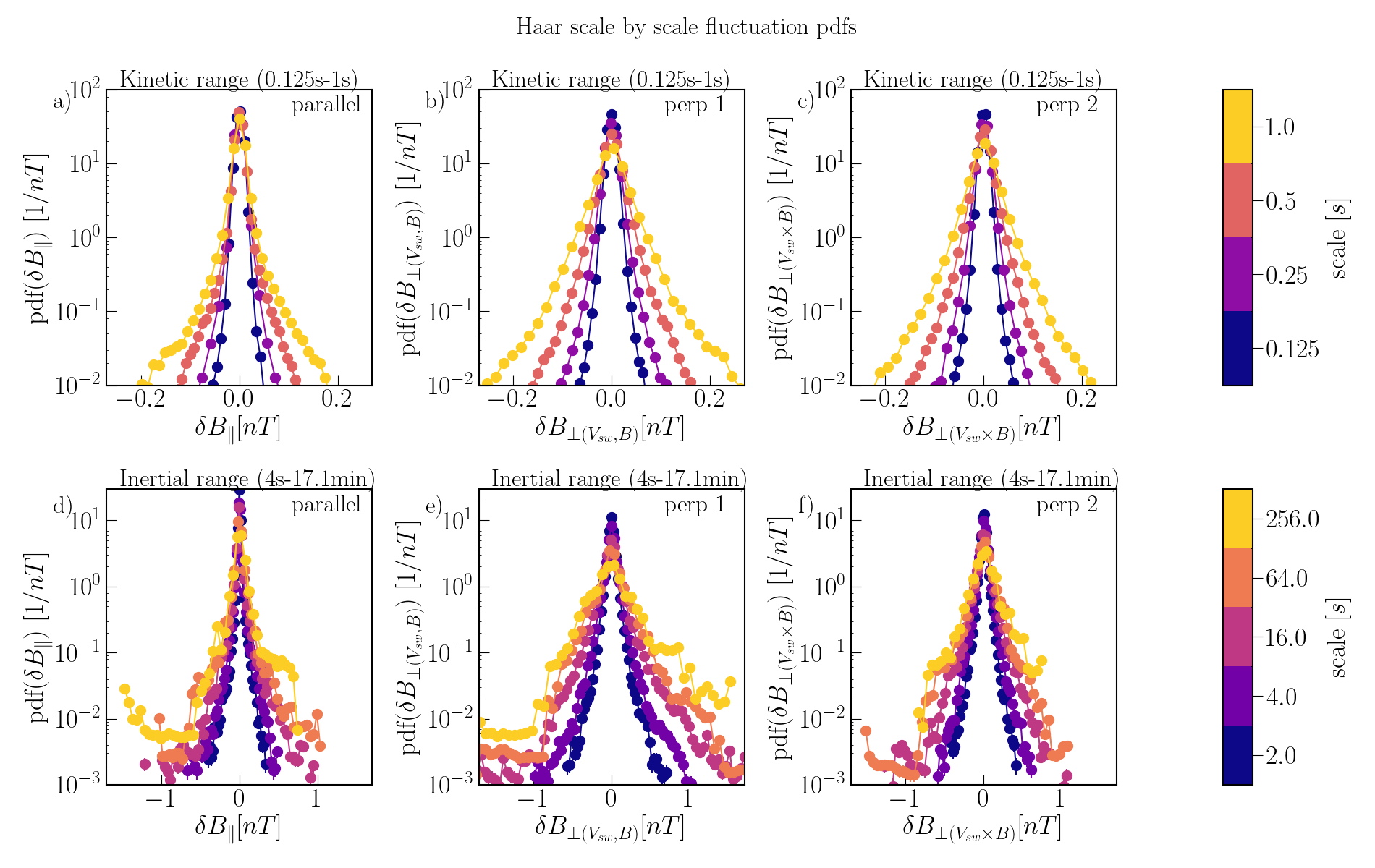}
    \caption{Probability distribution functions (pdfs) of Haar wavelet decompositions developing from stretched-tailed KR pdfs to Gaussian-like outer scale pdfs. Pdfs are shown for each component (columns) of the magnetic field for the KR (top row, panels a) to c)) up to $1$ s and IR (bottom row, panels d) to f)) for the interval at $0.989$ au from 2022-01-04. The colour marks the scale with largest scale in yellow and smallest scale in blue. The pdfs are normalised by bin width and overall number of samples of magnetic field data. The number of bins is scaled by the standard deviation $\sigma$ at the corresponding scale and bins with less than $10$ counts are discarded. The error is estimated as $\sqrt{n}$, where $n$ is the bin count, error bars are too small to be resolved visually.}
    \label{fig:pdfs_full_in1_h}
\end{figure}

\begin{figure}[ht]
    \centering
    \includegraphics[width=0.9\linewidth]{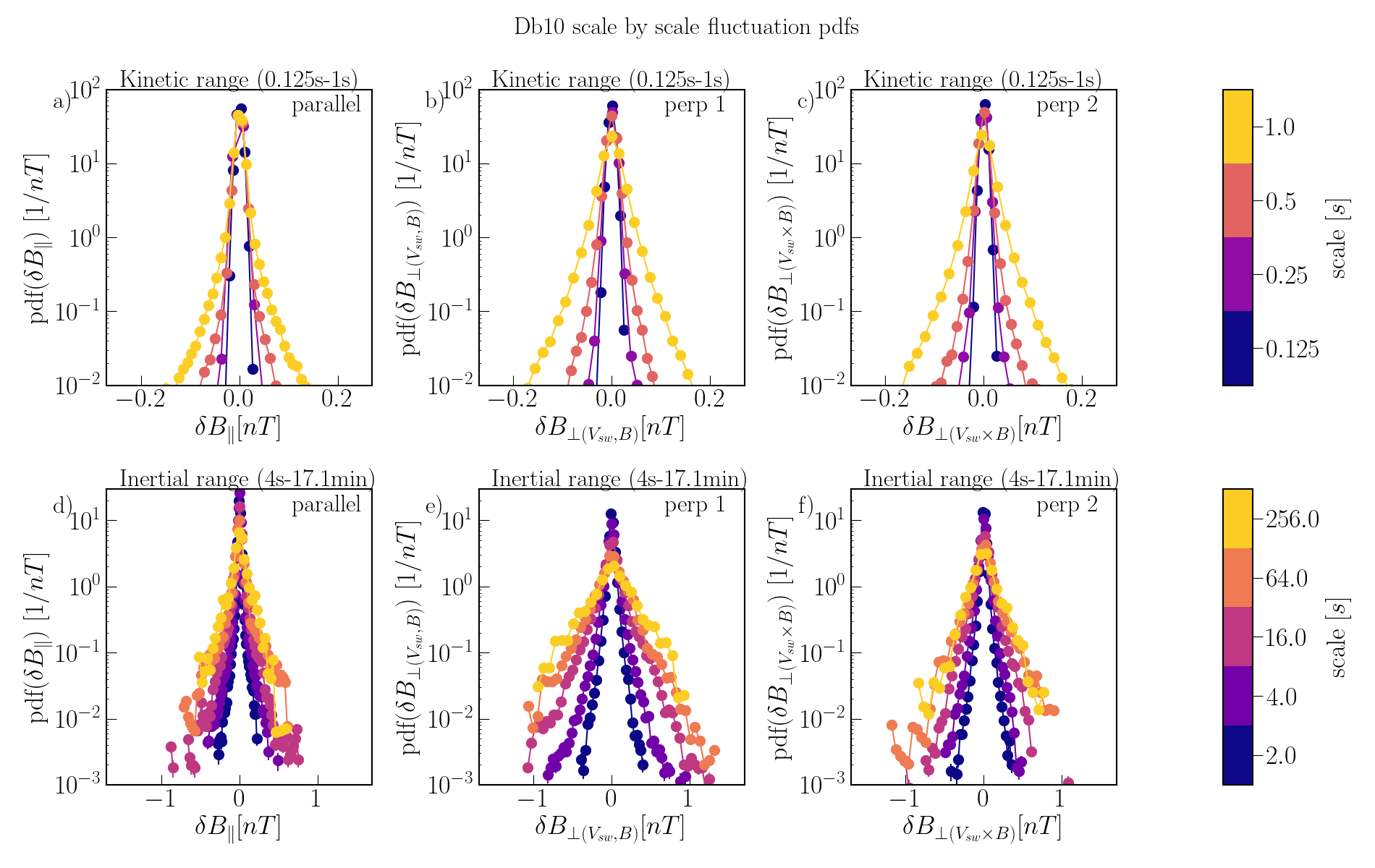}
    \caption{Probability distribution functions (pdfs) of Db10 wavelet decompositions developing from stretched-tailed KR pdfs to Gaussian-like outer scale pdfs. Pdfs are shown for each component (columns) of the magnetic field for the KR (top row, panels a) to c)) up to $1$ s and IR (bottom row, panels d) to f)) for the interval at $0.989$ au from 2022-01-04. The colour marks the scale with largest scale in yellow and smallest scale in blue. The pdfs are normalised by bin width and overall number of samples of magnetic field data. The number of bins is scaled by the standard deviation $\sigma$ at the corresponding scale and bins with less than $10$ counts are discarded. The error is estimated as $\sqrt{n}$, where $n$ is the bin count, error bars are too small to be resolved visually.}
    \label{fig:pdfs_full_in1_d}
\end{figure}

Turbulence is routinely studied by decomposing the observed time-series into fluctuations on different temporal scales.
Here, we compare the fluctuation pdfs extracted by the Haar (comparable to differencing), and Db10 wavelets, which resolve discontinuities, and wave-packets, respectively, to discriminate between wave-packets and coherent structures phenomenology at different scales within the turbulence cascade.
As we move from the shortest to the longest scales, the fluctuation pdf evolves from a sharply peaked functional form with extended tails to Gaussian-like at the outer scale of the turbulence inertial range (Figure \ref{fig:pdfs_full_in1_h} presents the Haar wavelet pdfs and Figure \ref{fig:pdfs_full_in1_d} the Db10 wavelet pdfs) \citep{bruno_probability_2004, alexandrova_small-scale_2008, frisch_turbulence_1995, tu_mhd_1995}.
The overall amplitude of the fluctuations, captured by their standard deviation, grows with temporal scale in a manner consistent with power-law scaling in the power spectral density (Figure \ref{fig:std} compares the standard deviation of the wavelet fluctuation pdfs for all intervals).
Specific coherent structures have been found to lie within the stretched tails of the fluctuation distributions (\cite{bruno_intermittency_2019} and references therein). Coherent structures  have been identified as origins of intermittency and sites of dissipation (e.g. \cite{osman_intermittency_2012, osman_kinetic_2012, greco_partial_2017, veltri_mhd_1999, gomes_origin_2022}). This confirms that the selected intervals are exhibiting the typical characteristics of turbulent fluctuations. 

\begin{figure}[ht]
    \centering
    \includegraphics[width=0.95\linewidth]{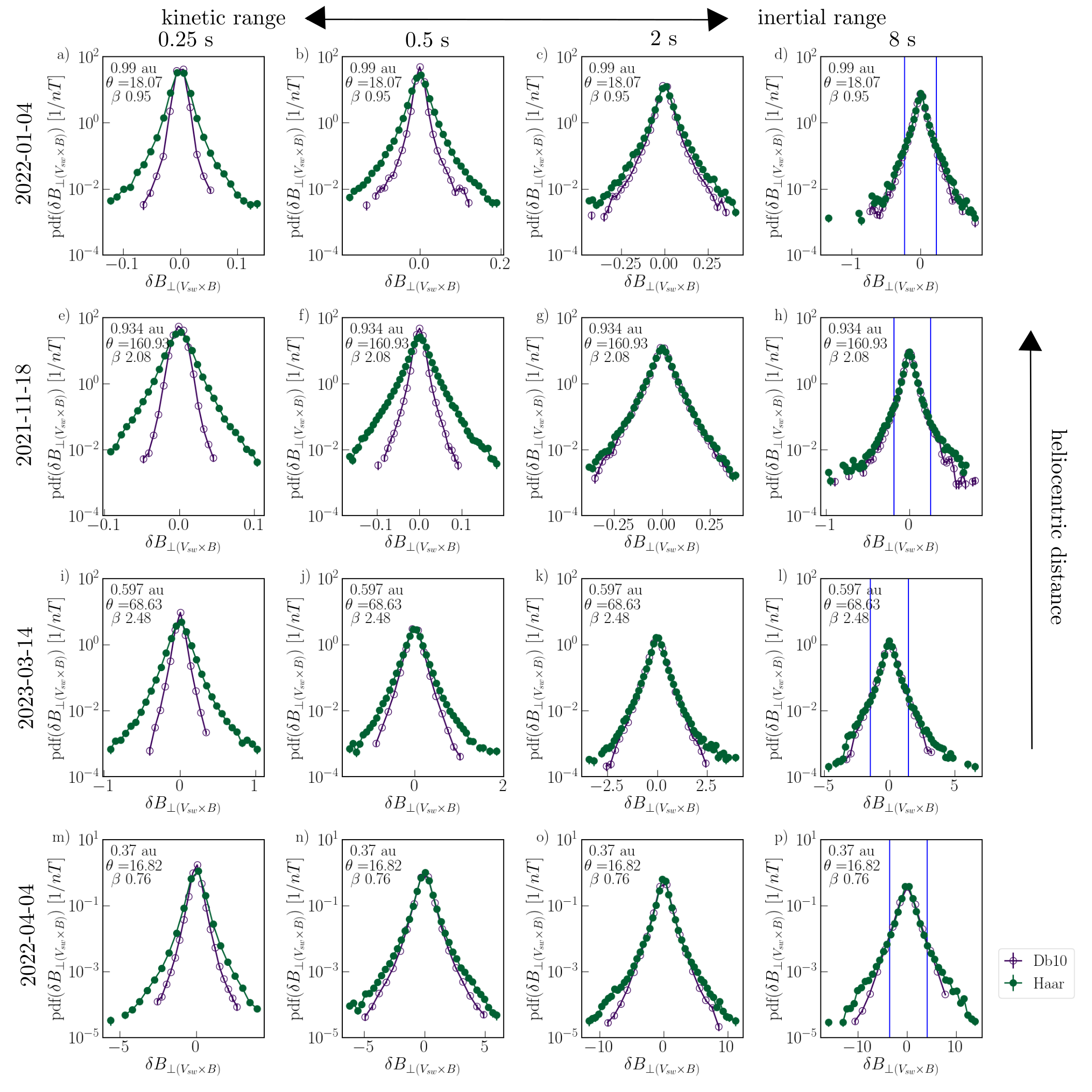}
    \caption{Probability distribution functions (pdfs) comparison between Haar and Db10 wavelet decompositions of the magnetic field. Pdfs are shown for $B_{\perp (V_{sw}\times B)}$ for four example intervals (rows). The chosen intervals (top down) are at $0.989$ au with $\beta=0.95$ and $\theta=18.07$°, at $0.934$ au with $\beta=2.08$ and $\theta=160.93$°, at $0.597$ au with $\beta=2.48$ and $\theta=68.63$°, and at $0.37$ au with $\beta=0.76$ and $\theta=16.82$°. The scales shown are increasing from left to right at $0.25$, $0.5$, $2$ and $8$ s. Empty purple circles are obtained from the Db10 wavelet decomposition, while green circles are from the Db10 wavelet decomposition. The pdfs are normalised by bin width and overall number of samples of magnetic field data. The number of bins is scaled by the standard deviation $\sigma$ at the corresponding scale and bins with less than $10$ counts are discarded. The error is estimated as $\sqrt{n}$, where $n$ is the bin count, error bars are too small to be resolved visually.}
    \label{fig:fluc_pdfs_perp2_zoom}
\end{figure}
In Figure \ref{fig:fluc_pdfs_perp2_zoom} we directly compare the fluctuation pdfs of the two wavelet decompositions across scales spanning the KR and IR. A full set of the fluctuation pdfs of both wavelet decompositions is provided in Figures \ref{fig:fluc_pdfs_para_in1}, \ref{fig:fluc_pdfs_perp1_in1} and \ref{fig:fluc_pdfs_perp2_in1} for each magnetic field component.
Four different intervals are shown in Figure \ref{fig:fluc_pdfs_perp2_zoom} (rows) where the heliocentric distance decreases from top to bottom.
The scales (columns in Figure \ref{fig:fluc_pdfs_perp2_zoom}) shown are at $0.25,~0.5,~2$ and $8$ s.
We find that three different morphologies of the pdfs can be seen in Figure \ref{fig:fluc_pdfs_perp2_zoom}.
In the KR (columns 1 and 2) the Haar (green circles) and Db10 (purple circles) fluctuation extremes, or tails, diverge. The Haar fluctuation pdf exhibits more stretched and extended tails than the Db10 wavelet decomposition.
Deep in the IR (column 4) there is a well-defined distribution core where the Haar and Db10 extracted fluctuation pdfs coincide. This core is between the blue vertical lines (column 4) whereas in the KR two distinct pdfs are found.
On intermediate scales (column 3) the pdfs have two components: The core of the fluctuation pdfs overlap, whereas the tails of the wavelet pdfs diverge. The tails of the pdf obtained from the Haar wavelet decomposition are more extended than those obtained from the Db10 wavelet decomposition.
The intermediate crossover range generally spans the spectral break scale obtained from the PSD.
This suggests three different regimes of turbulence: i) consistent with coherent structures in the kinetic range, ii) consistent with wave-packets deep in the inertial range, where the wavelet pdfs overlap and, iii) a crossover regime on intermediate scales, where a two component pdf is observed with tails consistent with coherent structures.

The distribution functions may differ either in their functional form, in their moments, or both. We can discriminate this with compensated Quantile-Quantile (QQ)-plots \citep{wilk_probability_1968, easton_multivariate_1990, tindale_solar_2017} of the wavelet fluctuation pdfs (see section \ref{sec:app-qq} for a description of Quantile-Quantile plots).
If the Haar and Db10 pdfs are drawn from the same distribution, then the compensated quantile trace will be a horizontal straight line at zero. If the pdfs are drawn from the same functional form but with different variance, the quantile trace will be a straight line diagonally.
A non-linear relationship on the QQ-plot indicates that the two distributions have different functional forms.
\begin{figure}[ht]
    \centering
    \includegraphics[width=0.9\linewidth]{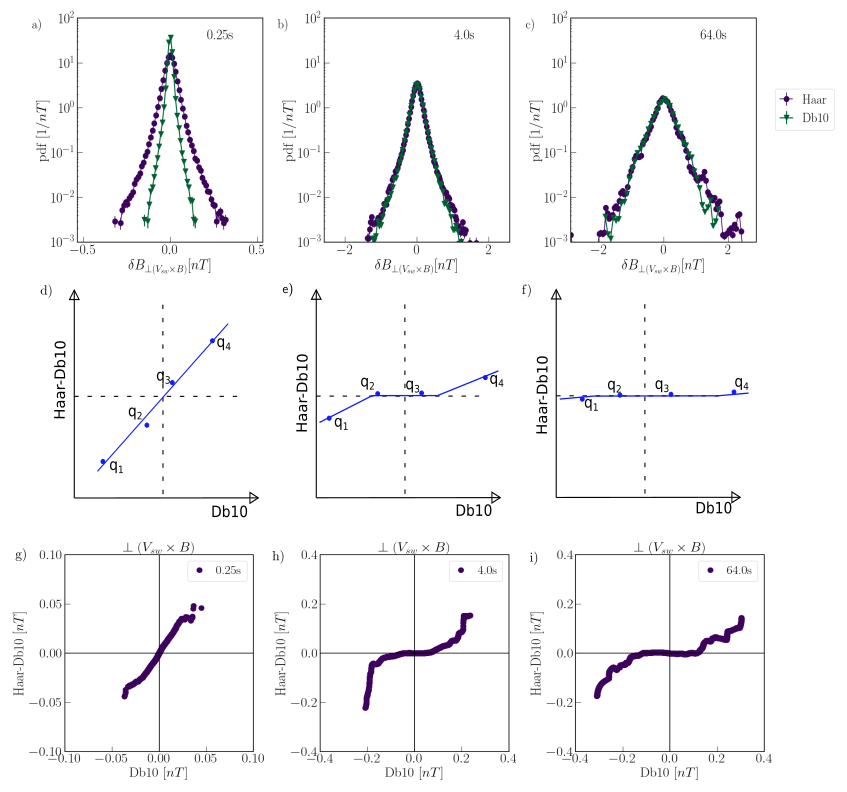}
    \caption{Compensated QQ-Plots compare the functional forms of the wavelet fluctuation pdfs. The compensated QQ-plots are of the $H-Db10$ versus the Db10 wavelet details for the magnetic field component $B_{\perp (V_{sw}\times B)}$. The distributions for three example scales in the three different regimes are presented with the corresponding compensated QQ-plots and their schematic drawing. The interval is at $0.9$ au from 2022-01-01. If the quantiles lie on the horizontal black line the distributions are the same.}
    \label{fig:qq-explained}
\end{figure}

\begin{figure}[ht]
    \centering
    \includegraphics[width=0.9\linewidth]{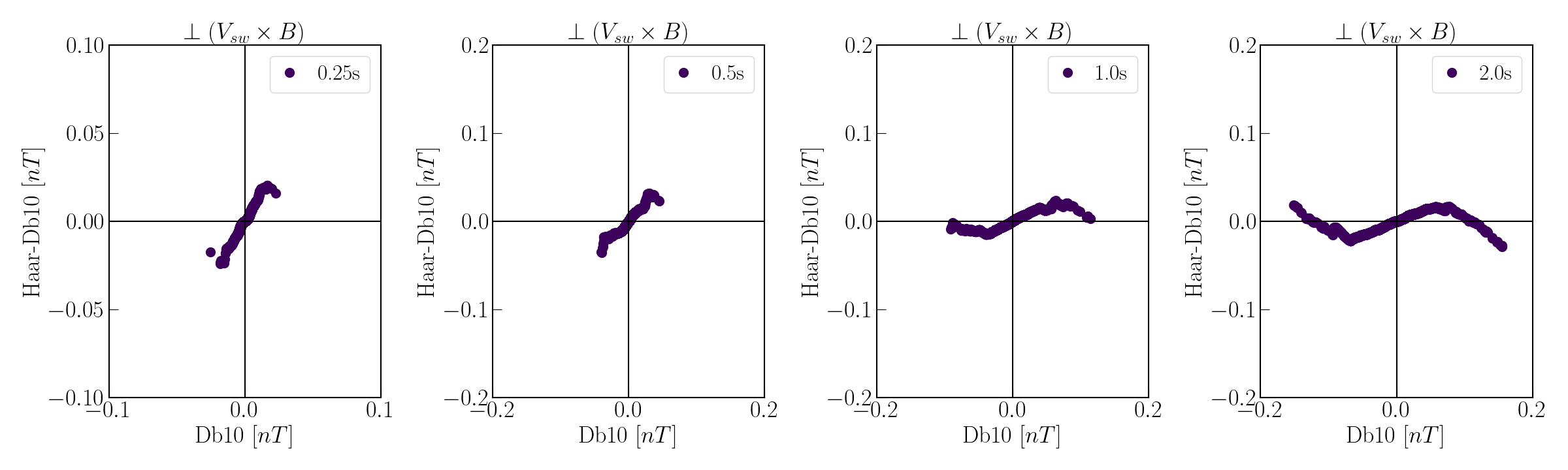}
    \caption{Compensated QQ-Plots compare the functional forms of the wavelet fluctuation pdfs. The compensated QQ-plots are of the $H-Db10$ versus the Db10 wavelet details for $0.25$ to $2$ s scales in the kinetic range. The interval is at $0.9$ au from 2022-01-04. If the quantiles lie on the horizontal black line the distributions are the same.}
    \label{fig:qq-kr}
\end{figure}

An example of analysis by QQ plot for this data is provided by Figure \ref{fig:qq-explained}. The Figure plots fluctuation pdfs in the KR (column 1), IR (column 3) and intermediate scales (column 2). The top row of panels overlay the Haar and Db10 wavelet fluctuation pdfs and we can see that whilst these coincide within the IR (panel c) the Haar fluctuation distribution is much broader in the KR (panel a). A more detailed comparison of the pdfs is afforded by QQ plots as shown in schematic (second row) and as obtained from the data (third row).
We have normalised the wavelet fluctuations by the overall magnetic field magnitude of each interval.
In the KR scale ($0.25$ s (column 1, panel d) in Figure \ref{fig:qq-explained}) the quantiles can then be seen to lie on a single line along $y=Ax$. This indicates that that the Haar fluctuation pdf has a larger variance than the Db10 fluctuation pdf but the pdfs share the same functional form.
This difference in variance between the two pdfs is given by the gradient of the QQ plot trace, $A$. This gradient can be seen from Figure \ref{fig:qq-kr} to decrease with increasing temporal scale.
On average the variance obtained from the Haar fluctuation pdfs is larger than that obtained from the Db10 fluctuation pdfs at kinetic range scales. The larger variance of the Haar fluctuation pdfs compared to the Db10 fluctuation pdfs is seen in the bottom row of Figure \ref{fig:std} where we present the percentage difference between Haar wavelet and Db10 wavelet estimates of the standard deviation with scale. The percentage difference is between $20$ and $125\%$ in the KR, while it is between $0-20\%$ in the IR.
In the IR scale in Figure \ref{fig:qq-explained} ($4$ and $64$ s, column 2 \& 3) the quantile trace has a central region which lies along $y=0$ so that the Haar and Db10 wavelet decomposition pdfs are similar in this central core. The largest fluctuations depart from this and form a distinct tail; more large fluctuations are obtained by the Haar wavelet decomposition than from the Db10 wavelet decomposition. The IR distributions are thus of a 2-component character with a central core distribution, where the wavelet decompositions have the same functional form and variance, and tails of same underlying functional form with different variance where the Haar wavelet decomposition resolves larger amplitude fluctuations.
\begin{figure}[ht]
    \centering
    \includegraphics[width=0.9\linewidth]{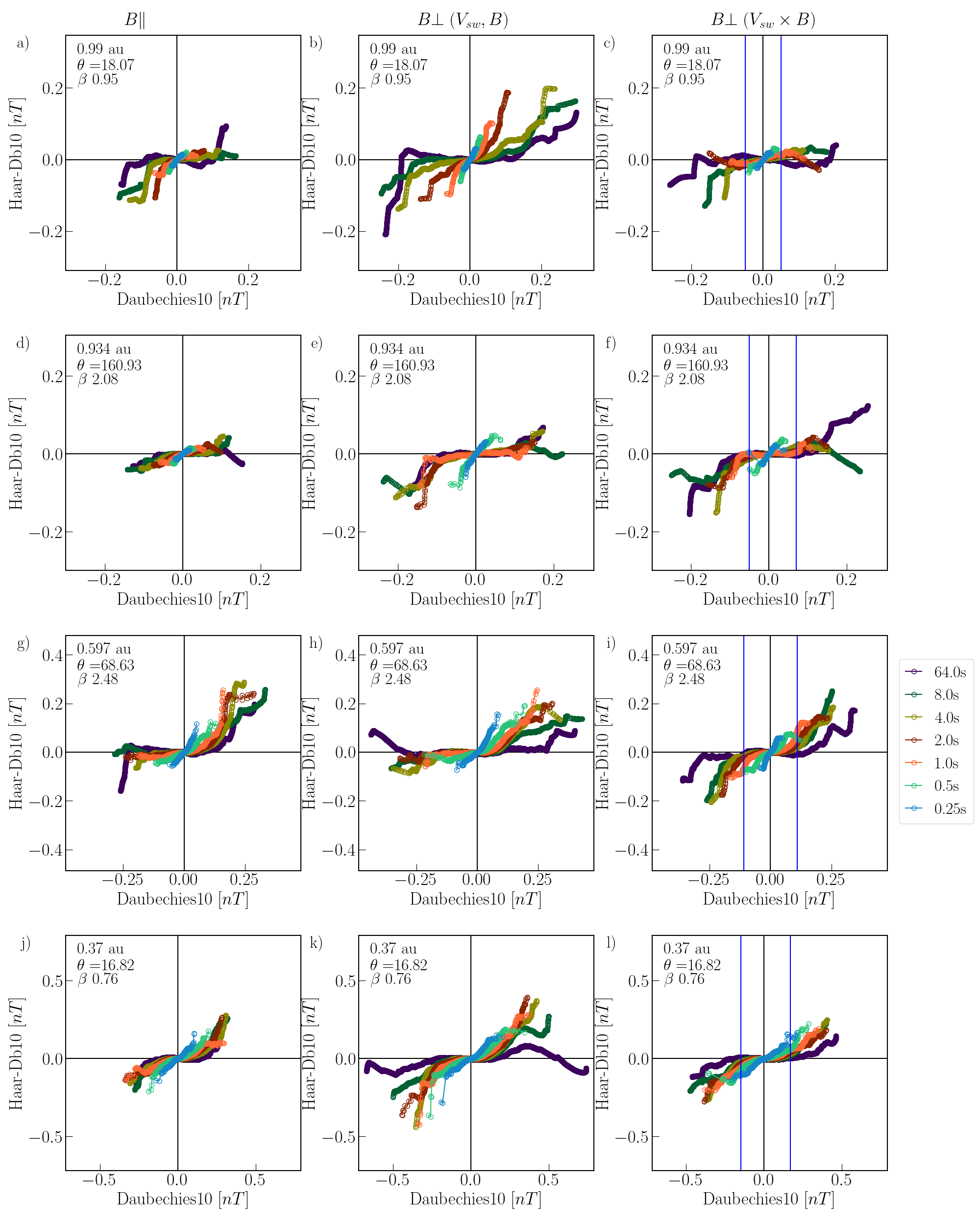}
    \caption{Compensated QQ-Plots compare the functional forms of the wavelet fluctuation pdfs. The compensated QQ-plots are of the $H-Db10$ versus the Db10 wavelet details over-plotted per scale for four intervals (rows) and all magnetic field components (columns). From top to bottom the intervals are at $0.9$ au, at $0.9$ au with $\theta=160.93$°, at $0.597$ au and $\beta=2.48$, and lastly at $0.3$ au. The different scales are denoted with different colours, the largest in purple, the smallest in bright blue. Scales from $0.25$ to $8$ s and additionally $64$ s scales are used. If the quantiles lie on the horizontal black line the distributions are the same.}
    \label{fig:qq-magnitude-cat}
\end{figure}
We present the compensated QQ-plots for four example intervals (Figure \ref{fig:qq-magnitude-cat}, rows) at $0.9$ au, at $0.9$ au with $\theta=160.93$°, at $0.597$ au and $\beta=2.48$, and lastly at $0.3$ au, and the magnetic field components (columns). Each colour refers to the fluctuations at a given temporal scale, the largest scale in purple at $64$ s and the smallest scale at $0.25$ s in teal. A full set of QQ-plots for all intervals is provided in Figure \ref{fig:qq-magnitude}. This figure shows the evolution of the fluctuation distributions from KR morphology, through the intermediate scales with 2-component character, to IR morphology.
Blue vertical lines (column 3) in Figure \ref{fig:qq-magnitude-cat} for $8$ s denote the limits of the core. These points are also marked in Figure \ref{fig:fluc_pdfs_perp2_zoom} (column 4). At $8$ s about $97$ \% of the fluctuations are within the core distribution between the blue lines. At $64$ s there is a small increase to an average of $98.5$ \%.
Within this overall behaviour there are differences depending on the heliocentric distance and field alignment angle $\theta$.
At $0.9$ au, $\beta=2.08$ and $\theta=160.93$° (panels e) to f)) the pdfs exhibits an abrupt crossover where at $1$ s (the spectral break) a core appears containing $97$ \% of the fluctuations, the fraction of fluctuations within the core does not increase with increasing scale. This abrupt crossover is not seen for the other high $\beta$ interval (panels h) and i)). A more comprehensive study may reveal other factors that affect the temporal scale and behaviour of the crossover.
For intervals $R\leq 0.4$ au a core is seen at $0.5$ s containing about $92$ \% of fluctuations in the core. The crossover range ends at $8$ s for $R\leq0.6$ au and at $16$ s for $R\sim0.9$ au. The crossover range is thus broader at small distances from the sun than at larger distances.

The first column in Figure \ref{fig:qq-magnitude-cat} shows the $B_{\parallel}$ component with the KR scales consistently as single line where the Haar wavelet fluctuation pdfs has larger amplitude tails and with increasing scale the core expands and the amplitude of the tails of the Haar wavelet fluctuation pdf decreases.
At $0.597$ au and large $\beta$ (row 3) the distributions show a mixture of behaviours, with $B_{\perp (V_{sw}\times B)}$ exhibiting the same evolution as intervals close to the sun, and $B_{\perp (V_{sw}, B)}$ like intervals at larger distances.

In summary, given that the Haar wavelet decomposition preferentially resolves coherent structures when compared to the Db10 wavelet, these results show that the KR is dominated by coherent structures across all amplitudes of fluctuations, whereas fluctuations in the IR are two component in character, with an extended tail dominated by coherent structures, and a core which can be be consistent with either coherent structures or wave-packets.

\section{Conclusions}\label{sec:conclude}
We performed scale-by-scale analysis of the magnetic field in a coordinate system ordered by the direction of the global, time-averaged background magnetic field for each of nine intervals of solar wind turbulence seen by SO for different plasma parameters and solar distances.
We compared time-series decompositions using the Haar (equivalent to differencing) and Db10 wavelets, which distinguish discontinuities (coherent structure phenomenology) and wave-packets (the phenomenology of wave-wave interactions) respectively. This work presents the first systematic comparison of these methods in the context of solar wind turbulence using wavelet decompositions that specifically characterize wave-like and coherent structure-like features in the time-series.
As we move from the shortest to the longest scales, the fluctuation pdf moves from a sharply peaked functional form with extended, super-exponential tails, to Gaussian at the outer scale of the turbulence \citep{frisch_turbulence_1995, camussi_orthonormal_1997}. The overall amplitude of the fluctuations, captured by their standard deviation, grows with temporal scale in a manner consistent with power-law scaling in the power spectral density (Figure \ref{fig:std}).
However we find that the fluctuation pdf functional form depends upon the decomposition used to obtain the fluctuations. We directly compared the pdfs of fluctuations obtained from Haar and Db10 wavelet decompositions.
We find that the fluctuation pdfs reveal three distinct morphologies
\begin{itemize}[nosep]
    \item Deep in the KR, the Haar and Db10 wavelet decompositions share the same functional form, but the the Haar wavelet decompositions have a variance that is larger than that obtained by the Db10, consistent with the phenomenology of coherent structures. 
    \item Deep in the IR there is a well-defined distribution core where the Haar and Db10 wavelet decomposition pdfs coincide and have the same functional form. The core contains about $98$ \% of fluctuations from the $64$ s scale.
    \item At intermediate scales between the IR and KR, the the Haar wavelet decomposition forms a larger amplitude pdf tail compared to that of the the Db10 wavelet decomposition. This is consistent with fluctuations in the distribution tails being dominated by coherent structures. 
    \item The intermediate crossover range of scales is located around the IR-KR spectral break scale. The characteristics of this crossover range depend on heliocentric distance.
    At distances around $0.9$ au the crossover range is quite narrow, from $4-16$ s. At around $0.3$ au the crossover occurs over a broader range of scales from $0.5-8$ s.
    \item  For one case where the field alignment angle $\theta$ is large, $160.93$°, the crossover is abrupt at $1$ s compared to other cases examined here. For this case, the fluctuation pdfs derived from the Haar and Db10 wavelet decompositions do not fully coincide even at the largest scales of the IR. This suggests further work to reveal which factors affect the character and temporal scale of the crossover.
\end{itemize}
Our results highlight the multi-component nature of the pdfs of fluctuations which can arise from either of two distinct phenomenologies that mediate the turbulent cascade, that of wave-packets, and coherent structures. We thus find that the fluctuation pdfs in the KR are consistent with coherent structure phenomenology. Deep in the inertial range the fluctuations pdfs of both wavelet decompositions coincide, which is consistent with either coherent structure or wave-packet phenomenology. On intermediate scales where we find a two-component pdf, the coherent structures dominate the pdf tails.

Formally, intermittency is a consequence of multifractality of the time-series \citep{frisch_turbulence_1995}. This corresponds to both non-Gaussian stretched tail pdfs and a scaling exponent of the structure functions $\zeta(q)$, defined by $S_{q}(\tau)=\langle |\delta B(\tau, t_{j})|^{q} \rangle_{t} \sim \tau^{\zeta(q)}$, which is non-linearly dependent on $q$ (monotonic curvature) \citep{kiyani_global_2009, kiyani_extracting_2006, frisch_turbulence_1995}. However, the presence of coherent structures does not require multifractality but it does imply non-Gaussian stretched tail pdfs.
In the kinetic range there is mono-scaling, that is $\zeta(q)$ is linear with $q$ and non-Gaussian stretched tail pdfs \citep{kiyani_global_2009, frisch_turbulence_1995}. Therefore, a picture of the phenomenology of solar wind turbulence is emerging in which there is multi-fractal scaling in the inertial range and fluctuations consistent with wave-packets, and mono-scaling in the kinetic range and fluctuations consistent with an enhancement of coherent structures. Our results systematically identify a crossover range in which there is a transition from multifractal to monofractal scaling via a $2$-component fluctuation distribution. Individual event studies based on PVI have also identified a sub-range of the high frequency IR \citep{wu_intermittency_2023}.

Additionally, we confirm previously reported results that
the IR-KR spectral break typically moves with the larger of the $\rho_{i}$ and $d_{i}$ scales depending on distance from the sun and $\beta$ \citep{bruno_radial_2014, lotz_radial_2023, safrankova_evolution_2023, chen_ion-scale_2014}.
The power in all components increases with decreasing distance from the sun \citep{chen_evolution_2020}.
We find that in the KR the two wavelet estimates differ, since the Haar wavelet cannot capture exponents steeper than $-3$ \citep{cho_simulations_2009, farge_wavelet_1991}.

In this paper we demonstrate how the Haar and Db10 wavelets resolve different underlying physics. Using the Haar and Db10 wavelets, we have detected a crossover from coherent structure phenomenology in the KR to wave-packet phenomenology in the IR. The crossover behaviour and range of scales depends on the heliocentric distance and field alignment angle. The population of coherent structures at small scales might suggest an association with the dissipation mechanism of turbulence, as suggested by the enhanced heating signatures found near coherent structures (e.g. \cite{osman_intermittency_2012, sioulas_statistical_2022}). A narrower crossover range of scales at large heliocentric distances may be connected to how well the turbulent cascade is developed. The larger range of coherent structures phenomenology at large distances may also be related to the evolution of intermittency with heliocentric distance (e.g. \cite{sioulas_magnetic_2022, bruno_radial_2003, pagel_radial_2003}).

This study only included one interval at large $\theta$ and one interval at $0.6$ au, which thus may only present outliers. A larger number of intervals at large $\theta$ as well as intervals at a variety of distances from the sun should be included in future work. An investigation of the coherent structures and waves present in the respectively dominated scales should give more insight into the physics present and how they connect to each other.

\begin{acknowledgments}
All data used in this study is freely available from the following sources (accessed last on \textbf{24th October 2023}):\dataset[Solar Orbiter Archive]{http://soar.esac.esa.int/soar}.

Solar Orbiter is a mission of international cooperation between ESA and NASA, operated by ESA.

This work is supported by funding for A.Bendt from the Science and Technologies Facilities Council (STFC). A.Bendt also acknowledges support from ISSI.
S.C.Chapman acknowledges support from ISSI via the J.Geiss fellowship and AFOSR grant FA8655-22-1-7056 and STFC grant ST/T000252/1.
T.Dudok de Wit acknowledges support from CNES.
\end{acknowledgments}

\software{MatLab \citep{noauthor_wavelet_2022}, Scipy \citep{virtanen_scipy_2020}, Numpy version 1.23.0 \citep{harris_array_2020}, Matplotlib version 3.7.1 \citep{hunter_matplotlib_2007}, Astropy \citep{robitaille_astropy_2013, collaboration_astropy_2018, collaboration_astropy_2022}, Cdflib \citep{stansby_mavensdccdflib_2022}}

\appendix

\section{Appendix}
\renewcommand\thefigure{\thesection.\arabic{figure}}
\setcounter{figure}{0}
\subsection{Supporting Methods}
\subsubsection{Quantile-Quantile Plots} \label{sec:app-qq}
\begin{figure}[ht]
    \centering
    \includegraphics[width=0.9\linewidth]{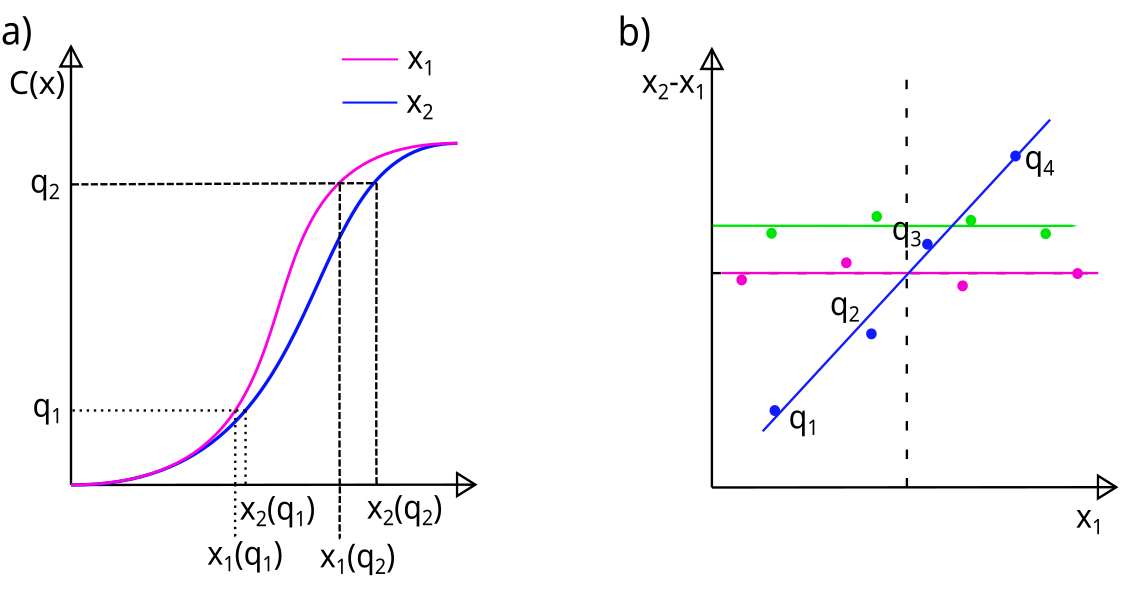}
    \caption{Diagrams showing the construction of the compensated QQ-plot. (a) The empirical CDFs of the samples $x_{1}$ and $x_{2}$. Proportion $q_{1}$ of the data set is bounded by quantile $x_{1}(q_{1})$ in sample $x_{1}$ and quantile $x_{2}(q_{1})$ in sample $x_{2}$, similar for $q_{2}$. (b) The compensated QQ-plot is produced by plotting $x_{1}(q)$ against $x_{2}(q)$ for all values of $q$. Pink line: $x_{2}-x_{1}=0$ i.e. are the same distribution, green line: $x_{2}-x_{1}=c$ i.e. different mean, blue line: $x_{2}-x_{1}=x_{1}$, i.e. different $\sigma$.}
    \label{fig:qq}
\end{figure}

\begin{figure}[ht]
    \centering
    \includegraphics[width=0.95\linewidth]{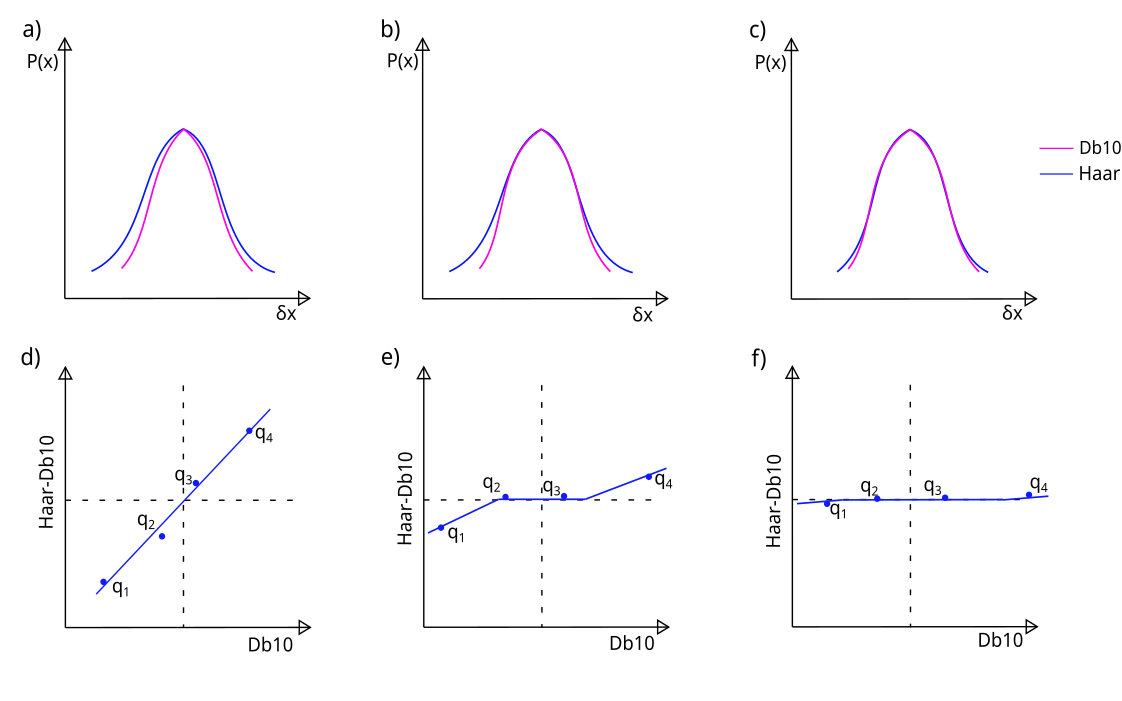}
    \caption{Diagrams showing the construction of the compensated QQ-plot. (a-c) The pdfs of the samples obtained from Db10 (pink) and Haar (blue) wavelets. (d-f) The compensated QQ-plot of the above pdf.}
    \label{fig:qq-regimes}
\end{figure}

Two distribution functions may differ in their functional form, or in their moments, or both. This difference can be seen in Quantile-Quantile (QQ)-plots \citep{wilk_probability_1968, easton_multivariate_1990, tindale_solar_2017}. These QQ-plots are constructed as follows (also see \citep{wilk_probability_1968, easton_multivariate_1990, tindale_solar_2017}). The cumulative density function $C(x)$ gives the likelihood of observing a value of $X \leq x$ as a function of $x$. The cdf takes values between zero and $1$ and defines the quantiles $x(q)$ of the distribution, so that $C(x(q))=0.5$ at the value $x(q)$ where $q=0.5$, the $0.5$ quantile, $C(x(q))=0.9$ at the value $x(q)$ where $q=0.9$, the $0.9$ quantile and so on.
The cdf is inverted to give the quantile function $x(q)=C^{-1}(q)$.
The QQ plot then compares the quantile functions of a pair of distributions $C_1$ and $C_2$ by plotting $x_1$ versus $x_2$ with the quantile $q$ as the parametric coordinate. The resulting QQ-plot has the values of the quantiles of $X$ on the axes of the two distributions to be compared, and the likelihood $q$ as parametric coordinate.
This is illustrated in Figure \ref{fig:qq} with two cdfs in panel a) and a compensated QQ-plot in panel b), where the $x_{1}-x_{2}$ is plotted versus $x_{2}$ with the quantiles as parametric coordinate $q$.
With the same functional form, the resulting line of quantiles can take three different shapes:
i) if $x_1$ and $x_2$ are drawn from the same distribution, then the compensated QQ-plot will be a straight line of $x_{2}-x_{1}=0$. ii) if the distribution has a shift in the mean, then it will be a straight line $x_{2}-x_{1}=c$ shifted from zero by $c$ and iii) if there is a change in the variance, then the compensated QQ-plot will be a straight line at $x_{2}-x_{1}=x_{1}$.
If the relationship on the QQ-plot is non-linear, the underlying functional forms of the distributions are different.
We use the Statistics and Machine Learning Toolbox from MatLab \citep{noauthor_statistics_2022} to determine the quantiles.
In the case of the wavelet fluctuation pdfs, the distributions show three different regimes illustrated with corresponding compensated QQ-plots in Figure \ref{fig:qq-regimes}:
i) the Haar has extended "fatter" tails than the Db10 and the distributions thus differ in $\sigma$ (panels a and d),
ii) the distributions are drawn from the same distributions in the core, but diverge in the tails (panels b and e),
iii) the distributions are drawn from the same distributions (panels c and f).


\subsection{Supporting Figures} \label{sec:sup-figs}
\subsubsection{Power Spectral Measures}

\begin{figure}[ht]
    \centering
    \includegraphics[width=0.45\linewidth]{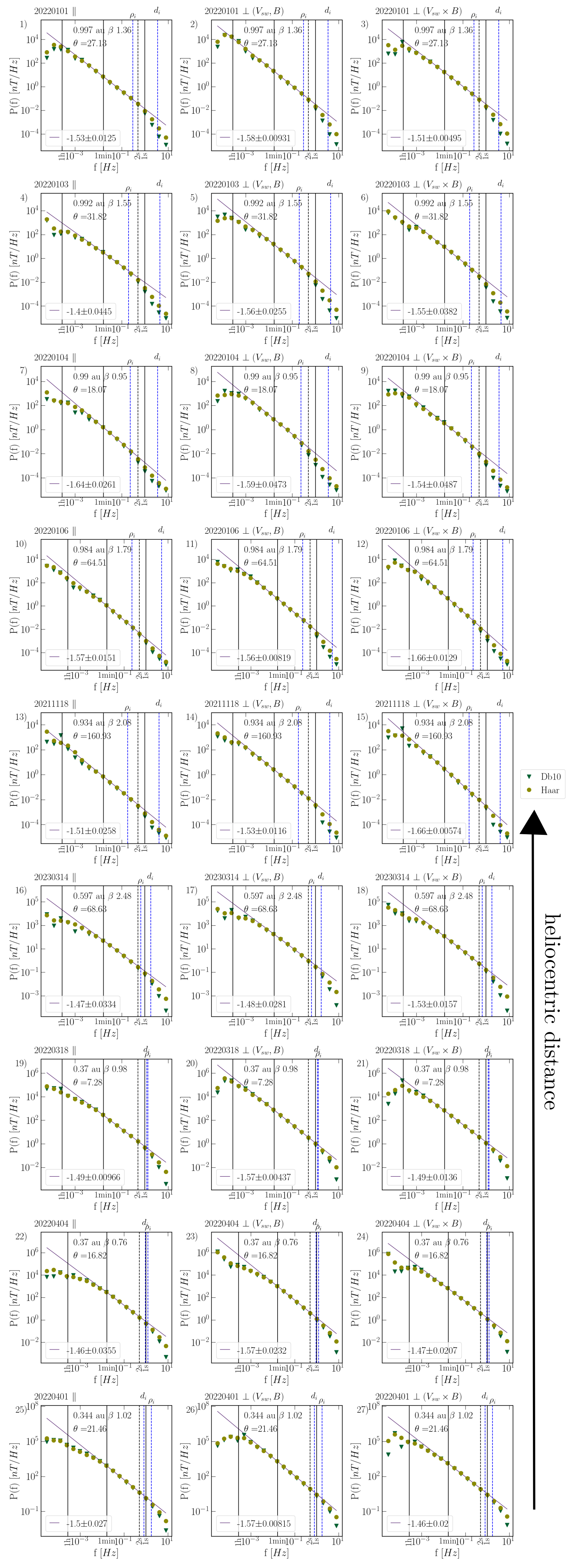}
    \caption{Power Spectra of all the different intervals (row) with decreasing distance from the sun and for each magnetic field component (column) and the respective scaling exponent in the inertial range fitted on the Haar wavelet. The 10th order Daubechies wavelet is shown by green triangles, yellow filled circles are the Haar wavelet. Blue vertical lines denote the ion-gyro frequency $\rho_{i}$  (dashed) and ion-inertia length $d_{i}$ (dash-dotted). Black vertical lines denote scales marked on the x-axis as $1$ s, $2$ s, $1$ min, $1$ h. Purple fit lines to the Haar wavelet power spectrum show the spectral exponent, which is quoted to three significant figures. Red crosses denote the power spectral estimate of the magnetic field magnitude $|B|$, and an orange line the linear fit to the $|B|$ power law.}
    \label{fig:psd-full}
\end{figure}

Figure \ref{fig:psd-full} presents the PSD for all intervals (rows) and each magnetic field component (columns). The increasing power levels are seen from top to bottom rows. The movement of $d_{i}$ and $\rho_{i}$ is seen clearly as a continuous shift from $\rho_{i}>d_{i}$ at $0.9$ au to $d_{i}>\rho_{i}$ at $0.3$ au. With the lower rows the lower KR break scale is seen as well as a smaller $1/f$-range break.

The second moment of the fluctuation pdfs relates to the PSD by definition and is an indicator for the overall power levels in the fluctuations for each component.
\begin{figure}[ht]
    \centering
    \includegraphics[width=0.95\linewidth]{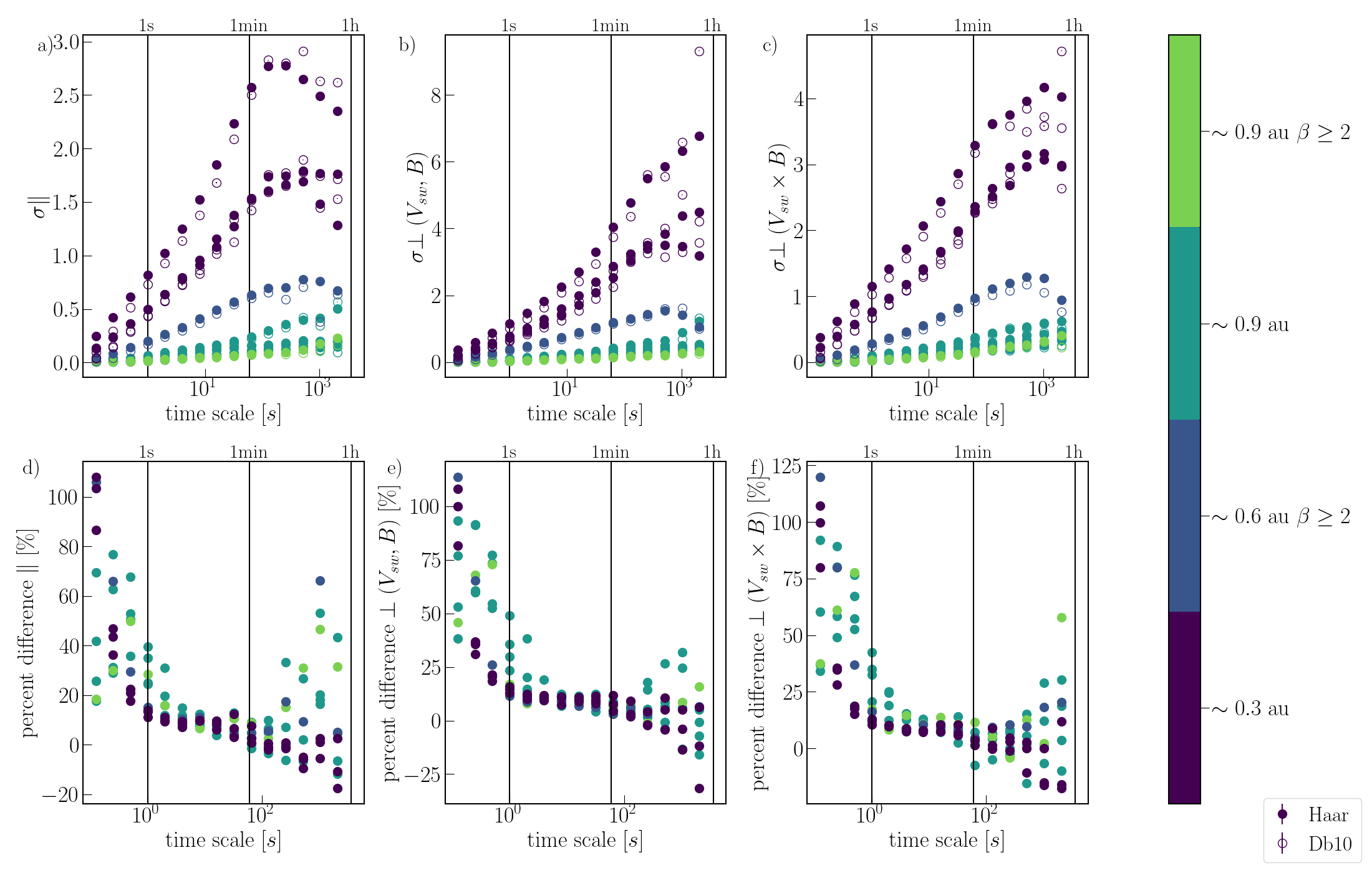}
    \caption{Comparison of standard deviation with increasing time scales for all intervals per magnetic field component. Open circles mark 10th order Daubechies wavelet, while filled circles are for the Haar wavelet of the corresponding colours per interval. The intervals are divided into four different intervals, with intervals at large distances in lighter colours. An early "roll-off" before the $1$ h scale is observed in $\sigma_{\parallel}$, for intervals close to the sun in $\sigma_{\perp (V_{sw}, B)}$ and for most intervals in $\sigma_{\perp (V_{sw}\times B)}$ in row 1. This is due to the early spectral break to the $1/f$-range in the power spectra. The error bars -too small to be seen- were determined by re-sampling and the variation of $\sigma$ values. In row 2 the percentage difference between Haar and Db10 wavelet estimates of $\sigma$ are presented.}
    \label{fig:std}
\end{figure}
Here we plot the standard deviation $\sigma$ of the fluctuation pdfs versus temporal scale in Figure \ref{fig:std} for all intervals.
As seen in the PSD (Figure \ref{fig:psd}), the Haar and Db10 wavelet generally agree on the standard deviation in the IR and only significantly diverge at large scales that move towards the upper end of the inertial range. The disagreement in the $1/f$-range is easily seen in the PSD Figure \ref{fig:psd} by an early "roll-off" into the $1/f$-range.
In terms of overall power there are three distinct groupings of these intervals. At $0.3$ au, the intervals show a progressively higher $\sigma$ compared to the intervals at $0.9$ au by a factor of $\sim 20$ at small scales, reducing to $\sim 6$ at larger scales. The magnetic field component $B_{\perp (V_{sw}, B)}$, has higher $\sigma$ values than any other component from about $100$ s and larger.

\subsubsection{Fluctuation distributions}
\begin{figure}[ht]
    \centering
    \includegraphics[width=0.9\linewidth]{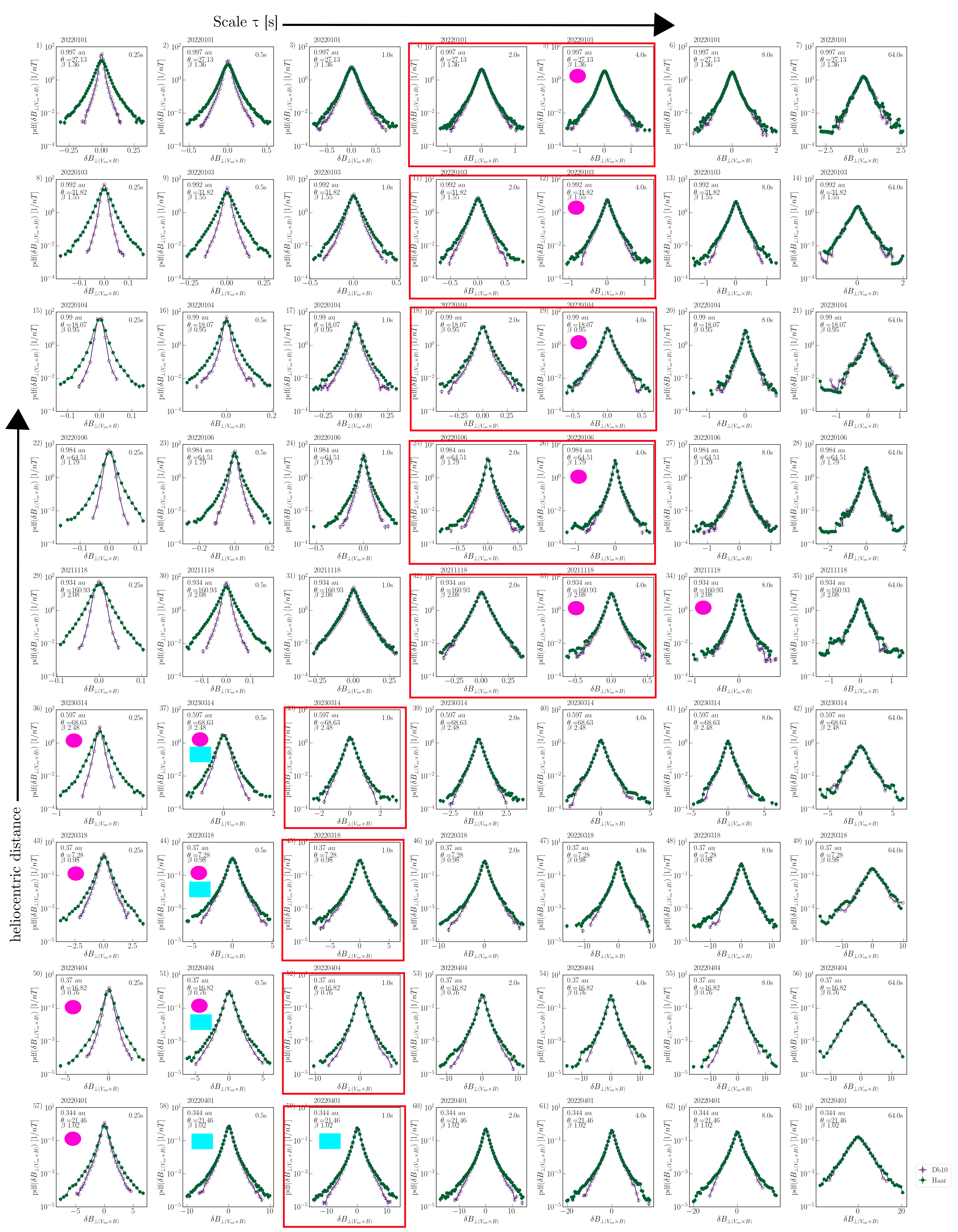}
    \caption{Probability distribution functions of wavelet fluctuations of $B_{\perp (V_{sw}\times B)}$ of all intervals in order of decreasing heliocentric distance (top to bottom) and increasing scale (left to right). Filled green circles are obtained from the Haar wavelet, while open purple circles are from the Db10 wavelet. The red box marks the spectral break scales. The pink circles denote $\rho_{i}$ and blue rectangles show $d_{i}$ (if two panels are marked the respective characteristic scale is between those two scales). The pdfs are normalised by bin width and overall number of samples of magnetic field data. The number of bins is scaled by the standard deviation $\sigma$ at the corresponding scale and bins with less than $10$ counts are discarded. The error is estimated as $\sqrt{n}$, where $n$ is the bin count, error bars are too small to be resolved visually.}
    \label{fig:fluc_pdfs_perp2_in1}
\end{figure}

The following Figures \ref{fig:fluc_pdfs_para_in1}, \ref{fig:fluc_pdfs_perp1_in1} and \ref{fig:fluc_pdfs_perp2_in1} show the fluctuation pdf comparison between Haar and Db10 wavelets for each interval (row) across scales from $0.25 -8$ and $64$ s (columns). The shift of $\rho_{i}$ (pink circles) and $d_{i}$ (blue rectangles) is seen, as well as the spectral break in red boxes.
The pdfs overlap largely in IR scales, and diverge in the tails in KR scales.

\begin{figure}[ht]
    \centering
    \includegraphics[width=0.9\linewidth]{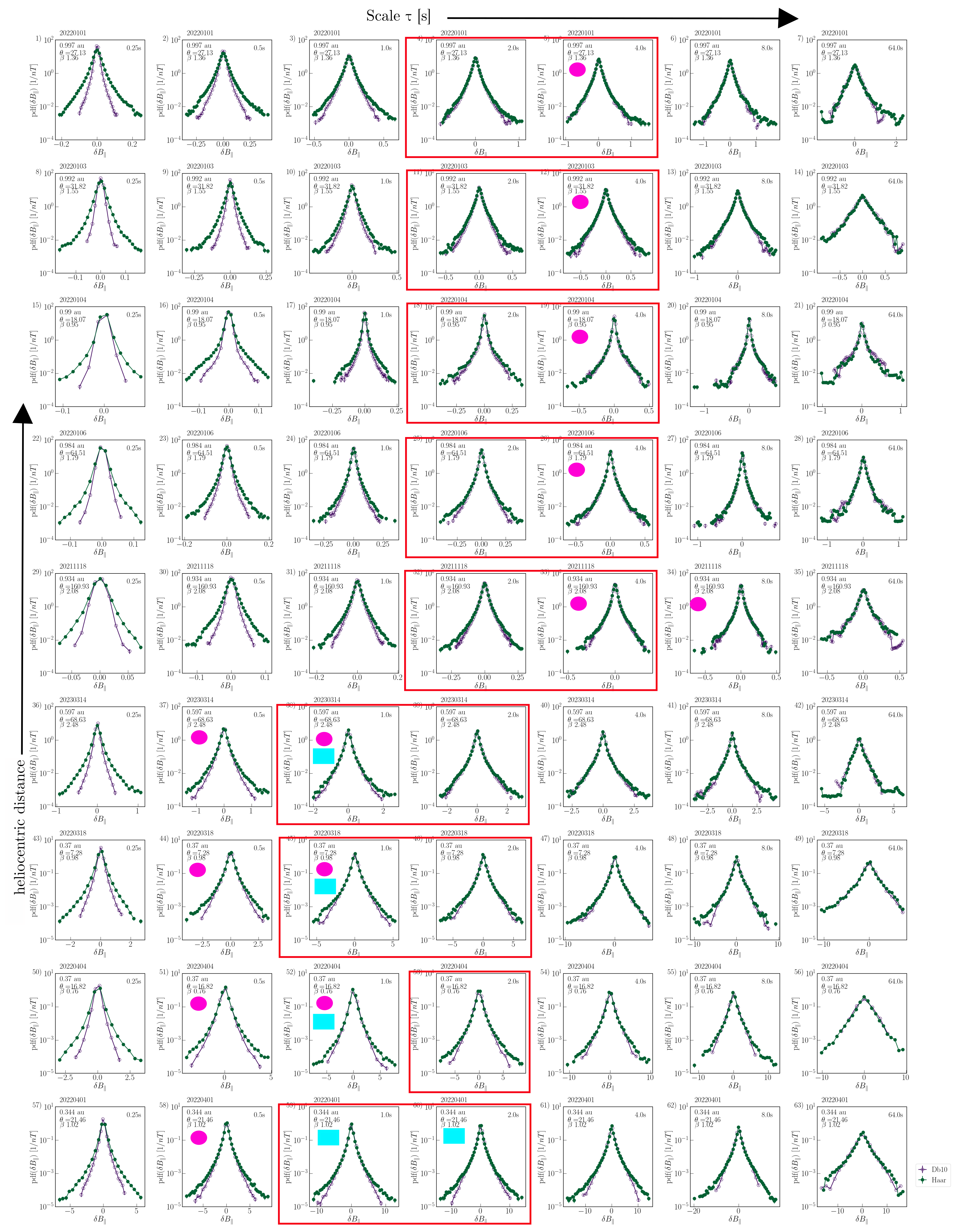}
    \caption{Probability distribution functions of wavelet fluctuations of $B_{\parallel}$ of all intervals in order of decreasing heliocentric distance (top to bottom) and increasing scale (left to right). Filled green circles are obtained from the Haar wavelet, while open purple circles are from the Db10 wavelet. The red box marks the spectral break scales. The pink circles denote $\rho_{i}$ and blue rectangles show $d_{i}$ (if two panels are marked the respective characteristic scale is between those two scales). The pdfs are normalised by bin width and overall number of samples of magnetic field data. The number of bins is scaled by the standard deviation $\sigma$ at the corresponding scale and bins with less than $10$ counts are discarded. The error is estimated as $\sqrt{n}$, where $n$ is the bin count, error bars are too small to be resolved visually.}
    \label{fig:fluc_pdfs_para_in1}
\end{figure}
\begin{figure}[ht]
    \centering
    \includegraphics[width=0.9\linewidth]{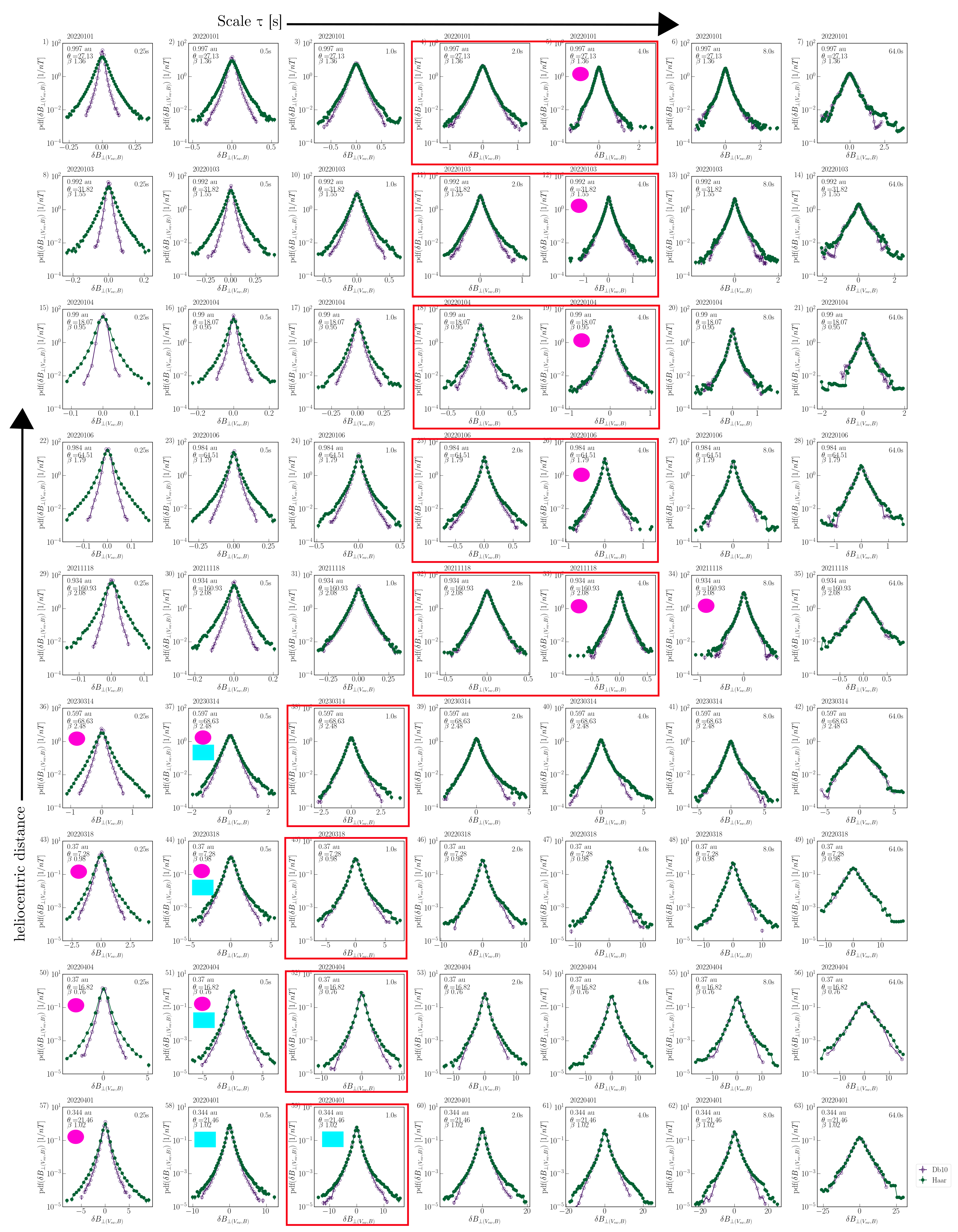}
    \caption{Probability distribution functions of wavelet fluctuations of $B_{\perp (V_{sw}, B)}$ of all intervals in order of decreasing heliocentric distance (top to bottom) and increasing scale (left to right). Filled green circles are obtained from the Haar wavelet, while open purple circles are from the Db10 wavelet. The red box marks the spectral break scales. The pink circles denote $\rho_{i}$ and blue rectangles show $d_{i}$ (if two panels are marked the respective characteristic scale is between those two scales). The pdfs are normalised by bin width and overall number of samples of magnetic field data. The number of bins is scaled by the standard deviation $\sigma$ at the corresponding scale and bins with less than $10$ counts are discarded. The error is estimated as $\sqrt{n}$, where $n$ is the bin count, error bars are too small to be resolved visually.}
    \label{fig:fluc_pdfs_perp1_in1}
\end{figure}

\begin{figure}
    \centering
    \includegraphics[width=0.45\linewidth]{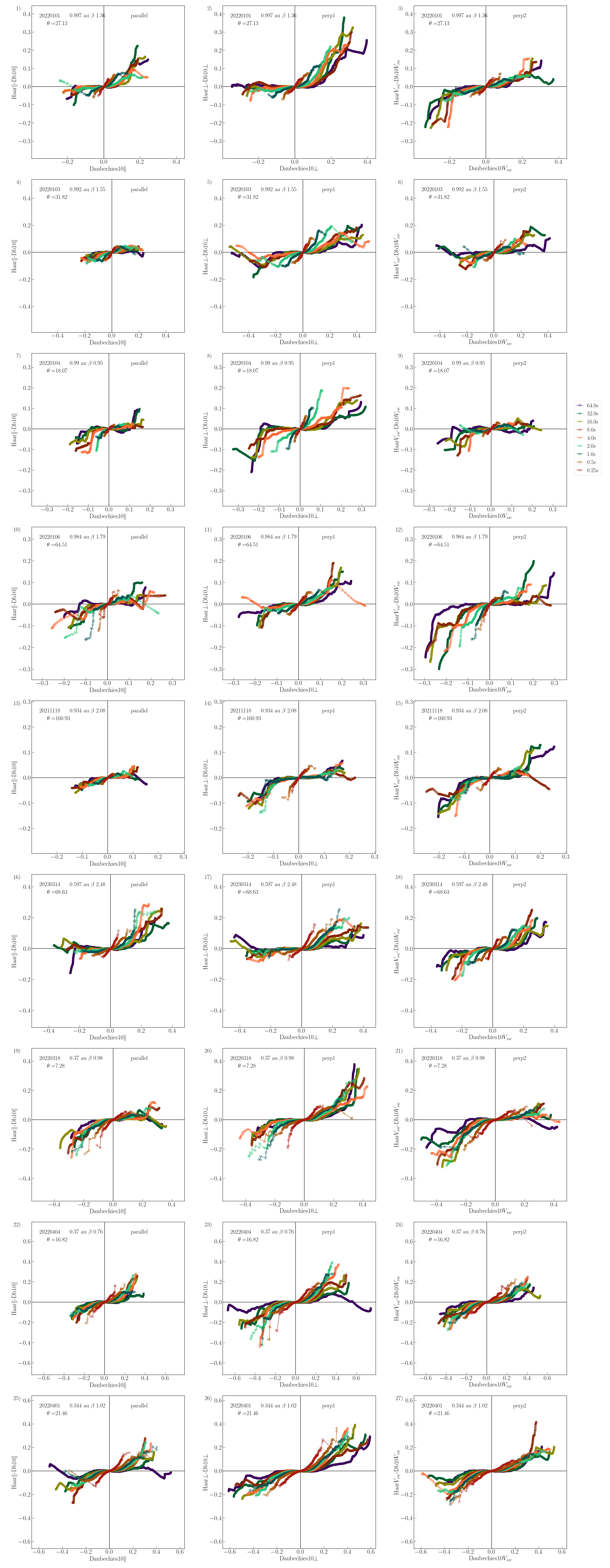}
    \caption{QQ-Plots of the Haar wavelet details versus the 10th order Daubechies wavelet details overlayed per scale for all intervals (rows, also labelled at the top right corner of the panels) and all magnetic field components (columns). The different scales are denoted with different colours. Scales from $0.25$ to $4$ s and additionally $64$ s scales are used. The fluctuations are normalised to the magnitude.}
    \label{fig:qq-magnitude}
\end{figure}
Figure \ref{fig:qq-magnitude} provides the compensated QQ-plots for all intervals (rows) for each magnetic field component (column). The gradual alignment of the cores is seen for all intervals and for intervals at $0.9$ au a single line with differing $\sigma$ for KR scales is visible, while intervals at smaller distances show an initial core in the KR pdfs. The tails are seen to decrease in slope with increasing scales.

\subsubsection{Time-series}
\begin{figure}[ht]
    \centering
    \includegraphics[width=0.9\linewidth]{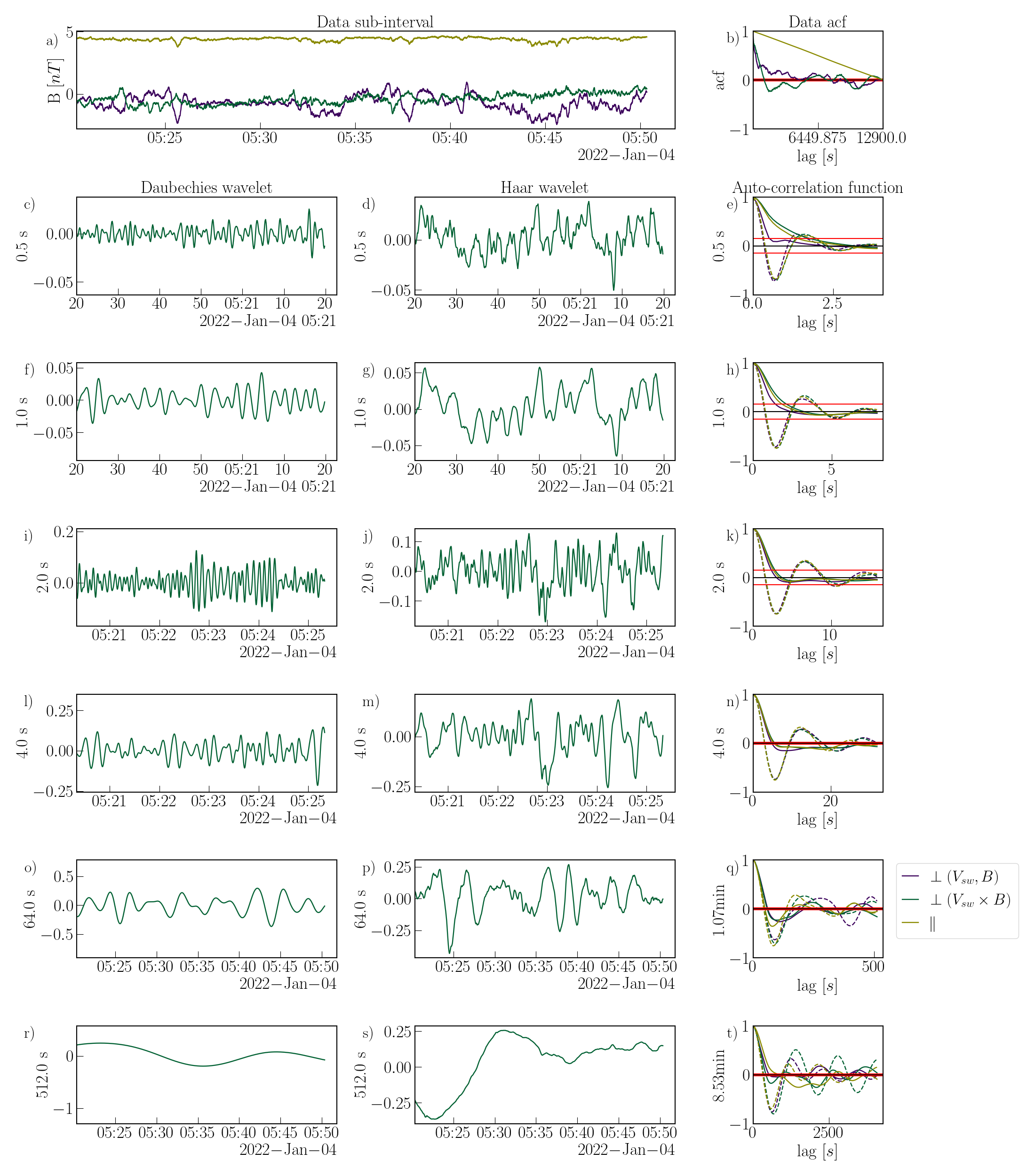}
    \caption{Decomposition of magnetic field time-series of interval at $0.989$ au from 2022-01-04 on scales $0.5$ to $4$ s and $64$ s and $512$ s of time-series (left) by Db10 and (middle) by Haar wavelet. (c) Haar wavelets. Right panels show acf of the middle third of decomposition time-series of Db10, dashed line and Haar, continuous lines. Red horizontal lines indicate the significance level obtained from auto-correlations of coloured noise with corresponding spectral exponents in the kinetic and inertial range. In purple the $B_{\perp (V_{sw}, B)}$, in green the $B_{\perp (V_{sw}\times B)}$, and finally in yellow $B_{\parallel}$.}
    \label{fig:dec_acf_full}
\end{figure}

Figure \ref{fig:dec_acf_full} displays the time-series sub-intervals and Haar and Db10 wavelet decompositions with corresponding acfs for the example interval 2022-01-04 at $\beta=0.95$ and $\theta=18.07$° for all magnetic field components.
With increasing scale, the fluctuations become more oscillatory and so does the acf. The Db10 continuously displays a more smooth and oscillatory time-series than the Haar wavelet.

\bibliography{references}{}
\bibliographystyle{aasjournal}

\end{document}